\documentclass{article}
\usepackage{arxiv}
\usepackage{graphicx}
\usepackage{cite}
\usepackage{amsmath}
\usepackage{url}
\usepackage{color}
\usepackage{multirow}
\usepackage{array}
\usepackage{subfig}
\usepackage{rays_defs_18}
\usepackage{balance}
\usepackage{booktabs}

\usepackage{algorithm}
\usepackage{algpseudocode}

\begin{document}
%

\title{Approximate Membership Query Filters with a False Positive Free Set}

\author{Pedro Reviriego, Alfonso Sánchez-Macián,  Stefan Walzer and Peter C. Dillinger}  


\thanks{Pedro Reviriego is with Universidad Carlos III de Madrid, Legan\'es 28911, Madrid, Spain. email:
        {\tt\small revirieg@it.uc3m.es}}%
\thanks{Alfonso Sánchez-Macián is with Universidad Antonio de Nebrija, 28040, Madrid, Spain. email: {\tt\small asanchep@nebrija.es}}%
\thanks{Stefan Walzer is with University of Cologne, Cologne, Germany. email:
        {\tt\small walzer@cs.uni-koeln.de}}%
\thanks{Peter C. Dillinger is with Facebook, Inc., Seattle, Washington, USA.
 email:  {\tt\small peterd@fb.com}}%

\maketitle


\begin{abstract}

In the last decade, significant efforts have been made to reduce the false positive rate of approximate membership checking structures. This has led to the development of new structures such as cuckoo filters and xor filters. Adaptive filters that can react to false positives as they occur to avoid them for future queries to the same elements have also been recently developed. In this paper, we propose a new type of static filters that completely avoid false positives for a given set of negative elements and show how they can be efficiently implemented using xor probing filters. Several constructions of these filters with a false positive free set are proposed that minimize the memory and speed overheads introduced by avoiding false positives. The proposed filters have been extensively evaluated to validate their functionality and show that in many cases both the memory and speed overheads are negligible. We also discuss several use cases to illustrate the potential benefits of the proposed filters in practical applications. 

\end{abstract}



\section{Introduction}

Approximate membership checking is widely used in many computing applications \cite{BFDS}. For example, to accelerate access to key-value stores \cite{MonkeyLSM} or to data stored in external memory \cite{EMOMA}. Approximate membership checking is also used in networking, for example for packet classification \cite{CFBF}. The most frequently used data structure for approximate membership queries is probably the Bloom filter \cite{BF} for which many enhancements and optimizations have been proposed \cite{BFs_Ori}. A Bloom filter maps an element to a set of bits in a bit vector using a group of hash functions and sets those bits to one on insertion and checks if they are one when checking membership. This construction ensures that there are no false negatives. Instead, false positives can occur when other elements have set to one the positions checked by a given element. Therefore, the approximate checking manifests as a false positive probability for elements not originally added to the filter \cite{BF}. This is the price paid for having a much smaller and faster representation of the set. Alternative approximate membership querying structures have also been proposed like the cuckoo filter \cite{CF} or more recently the xor filter \cite{XORfilter} that can achieve smaller false positive probability with the same memory budget in some settings.

Since false positives are the main drawback of all these approximate membership querying structures, significant efforts have been made to reduce them for example by adapting the filter to react to false positives as they occur \cite{Bender},\cite{ACF}. Another approach is to design filters that are false positive free when the set stored is small and from elements taken also from a small universe \cite{FPFZ}. Unfortunately, this limits the use of the false positive free filters to very small sets having only a few elements. In this paper we introduce a new type of filters that are false positive free on a subset of the negative elements. That is, the proposed filters do suffer false positives over the entire universe but we can select a set of negatives, that we assume can be larger than the set of elements stored in the filter, that will not suffer false positives. This is useful to reduce the observed false positive rate by including the most commonly accessed negative elements on that set. 

To better understand the problem we address in this paper, let us consider one of the initial applications of a Bloom filter, storing the set of correct words to check their spelling \cite{BFSpellCheck}. In more detail, all the words of the dictionary are stored in the Bloom filter and used to check written words so that an error is flagged when the Bloom filter returns a negative. This design ensures that valid words are always classified as correct but there may be some invalid words that are not detected. An interesting observation is that there are some common misspellings that occur frequently for which it would be beneficial not having false positives on the filter. However, the Bloom filter treats all the words that are not in the dictionary equally having all the same false positive probability. 

In this paper we consider this problem: given two disjoint sets $S$ and $T$ of elements that belong to universe $U$, design a false negative free filter that stores $S$ and has no false positives for elements in $T$ and a low false positive probability for the rest of the elements that are neither in $S$ nor in $T$. That is, our goal is to design a filter for which elements in $S$ are guaranteed to query positive, elements in $T$ are guaranteed to query negative, and most other elements query negative. We denote these filters are false positive free on a set ($T$) that is possibly larger than $S$. Obviously such a filter could be used with a set $S$ of the words in the dictionary and a set $T$ with the common misspellings to make the filter based spell checker more effective. These filters with a false positive free set would also be beneficial in applications such as filtering for database queries \cite{MonkeyLSM}, in memory key-value stores \cite{EMOMA}, for filtering of URLs\cite{URLblacklist} or in lightweight Bitcoin nodes \cite{privacyblockchainggervais2014} among many others as will be discussed in several case studies in section~\ref{Cases}. 

The rest of the paper is organized as follows, section \ref{Prel} covers the preliminaries by briefly reviewing existing approximate membership query filters, the efforts to reduce their false positive rate and providing an overview of xor probing filters that are used to build our proposed filters. The proposed filters are presented and analyzed in section~\ref{FPFS}. The evaluation of the filters is presented in section~\ref{Evaluation} that also compares them with existing filters. Section~\ref{Cases} discusses several use cases where the use of the proposed filters would be beneficial. Finally, the paper ends in section~\ref{Conclusions} with the conclusion and some ideas for future work.

\section{Preliminaries}
\label{Prel}

\subsection{Approximate membership query filters}

As discussed in the introduction, approximate membership checking is widely used in computing and networking and different structures have been proposed to implement it. The most well-known and commonly used option is Bloom filters \cite{BFs_Ori} but other alternatives like the Bloomier filter \cite{chazelle2004bloomier}, cuckoo filters \cite{CF} or xor filters have been proposed in the last two decades \cite{XORfilter}. In all cases, elements are mapped to several positions on an array or table using a set of hash functions $h_1(x),h_2(x),\ldots,h_k(x)$. The contents of those positions are used to check the membership. The filters can be classified according to how the checking is done \cite{Ribbonfilter}: 

\begin{itemize}
    \item \textit{And probing:} the logical \textit{And} of the positions is used to check membership. This is the case of the Bloom filter in which an array of bits is used and true is returned when the and of positions $h_1(x),h_2(x),\ldots,h_k(x)$ is a one and a false is returned otherwise.
    \item \textit{Or probing:} the logical \textit{Or} of the matching of the contents of the positions with a value $v(x)$ derived from the element is used to detect positives. This is the scheme used in the cuckoo filter that computes a fingerprint per element and compares it with the fingerprints stored in the positions and if any of them matches the element's fingerprint true is returned \cite{CFBF}.
    \item \textit{Xor probing:} the logical \textit{Xor} of the positions is used to compute a value $F_{xor}(x)$ that is used to detect positives. A  value derived from the element $v(x)$ is compared with $F_{xor}(x)$  and true is return on a match and false otherwise. This is the scheme used in the xor and ribbon filters \cite{XORfilter},\cite{Ribbonfilter}. 
\end{itemize}

Each filter type and construction has advantages and disadvantages. For example, Bloom filters are the simplest to construct and their false positive probability (FPP) degrades gracefully if the number of elements added does not match what was predicted and optimized for. However, they do not support deletions unless counters instead of bits are used and their FPP is large in some settings \cite{BFs_Ori}. Another advantage of Bloom filters is that they can be implemented on a single memory access \cite{BBF}, so can be made faster than cuckoo filters in software \cite{CFBFSpeed}. While cuckoo filters support removals and have lower FPP in many settings, space efficiency with low FPP depends on high occupancy but not so high to fail insertion \cite{CFBF}. Finally, xor filters \cite{XORfilter} and ribbon filters \cite{Ribbonfilter} achieve better FPP for the same memory budget in many cases, with good speed, but do not support dynamic insertions and removals, limiting their use to \emph{static} filtering applications, in which real time updates to the set are not needed. In summary, the best filter configuration depends on the application requirements and implementation platform.

\subsection{Reducing the false positive probability of filters}

First, when the universe of elements is sufficiently small, it is possible to represent an exact set with no false positives, using less space than a generic filter with low but non-zero FPP, e.g.~\cite{ClearyCompactHash}. It is even possible for exact sets to degrade into filters with non-zero FPP as more elements are added \cite{FPFZ,FPFZCMS,DillingerManolios2009}. We expect only a small minority of applications to have universes sufficiently small to justify these approaches. The rest of the paper assumes a large or infinite universe.

One of the main performance parameters of approximate query filters is the FPP that they achieve for a given memory budget and many efforts have been made to reduce it. For example, for Bloom filters optimizations like the Variable Increments Bloom filter have been proposed to reduce the FPP \cite{BF}. Indeed, one of the main motivations to develop cuckoo, xor and ribbon filters was to improve space efficiency in the FPP.

Another strategy is to reduce selectively the FPP only for elements that are frequently accessed. This can be done in the Bloom filter by setting some positions to zero in the array that correspond to false positive elements that are frequently accessed. This however introduces false negatives for other elements \cite{RetouchedBF}. The knowledge of elements that are frequently accessed may not be known in advanced, and in that scenario filters have to remove false positives adaptively as they occur \cite{Bender}. For example, Adaptive Cuckoo Filters (ACFs) change the stored fingerprints to remove false positives as they occur \cite{ACF}.  

As discussed in the introduction, in this paper we focus on the design of filters that are false positive free on a set of elements $T$ while allowing false positives for the rest of the negative elements in the universe. To the best of our knowledge efficient designs for such filters when $T$ is large have not been proposed before. False positives can be avoided by using a Bloomier filter \cite{chazelle2004bloomier} but each such false positive requires an entry on the filter thus incurring a large cost when $T$ is large. There have also been efforts to reduce the false positives for a given subset of frequently accessed elements like for example \cite{StackedBFs},\cite{ACF} but not to completely avoid them by construction.

\subsection{Xor probing filters}

An xor probing filter can be seen at a high level as a function that for a given element $x$ returns a value $F_{xor}(x)$ that is obtained by performing the xor of several positions on the filter given by hash functions $h_1(x),h_2(x),\ldots,h_k(x)$. The filter is constructed so that $F_{xor}(x) = v(x)$ holds when an element $x$ has been inserted into the filter, where $v(x)$ is a fingerprint computed on the element. Instead when the element has not been inserted on the filter, in most cases $F_{xor}(x) \ne v(x)$ and the probability of having $F_{xor}(x) = v(x)$ would be approximately $2^{-r}$, where $r$ is the number of bits in $v(x)$.

A lookup on an xor probing filter is a simple operation that just requires accessing positions  $h_1(x),h_2(x),\ldots,h_k(x)$, computing the xor of their values to obtain $F_{xor}(x)$ and comparing it with $v(x)$. On a match, true is returned and false is returned otherwise. The construction of xor probing filters is more complex and needs to be precomputed, so these filters are not suitable for applications that need to perform frequent updates on the stored elements. Two constructions have been proposed so far, the xor filter \cite{MWHC,XORfilter} and the ribbon filter \cite{Ribbonfilter}. The first one usually probes $k=3$ random positions while the second probes many more positions in a small window. A standard xor filter uses $1.23 \cdot r$ bits of memory per element, already a significant improvement on Bloom filter's best case of $1.44 \cdot r$. With some elaboration \cite{StaticFunctions1,SpatialCoupling}, both constructions approach information-theoretic limits, requiring as little as $1.01 \cdot r$ bits with negligible additive overheads. This makes them the most memory efficient approximate membership query structures to date for many settings. In the following, we assume that in general an $r$ bit xor probing filter requires $(1+\epsilon) \cdot r$ memory bits per element.

In the rest of the paper we will use xor probing filters to construct our filters with false positive free sets in a general form. The specifics of xor filter constructions will be covered in the evaluation section when analyzing the practical performance of our filters. Before presenting the proposed filters in the next section, the main notations used in the rest of the paper are summarized in Table \ref{table:notations}.

\def\FPP{\mathrm{FPP}} 
\begin{table*}[h]
\centering
\caption{Summary of main notations}
\begin{tabular}{cl}  
\toprule
\textbf{Symbol}  & \textbf{Meaning} \\ 
\midrule
$U$ & universe of elements \\
$S$ & set stored in the filter \\
$T$ & set for which false positives must be avoided \\
$F$ & residual set for which false positives must be avoided \\
$r$ & number of bits of the base xor filter \\
$c$ & number of subfilters of the second filter in the Integrated Filter (IF) design \\
$a$ & number of bits added to the first filter \\
$\FPP$ & false positive probability \\
$\epsilon$ & memory overhead required by the xor probing filter \\
\bottomrule
\end{tabular}
\label{table:notations}
\end{table*} 
 
\section{Avoiding False Positives for a given set}
\label{FPFS}

As discussed in the introduction, our goal is to design filters that provide approximate membership checking for a set $S$ ensuring that all elements of set $T$ return a negative on the filter. In the following subsections we first formally define the target filters, analyze their lower bound memory requirements theoretically and discuss briefly several potential options to construct the filters. Then, the proposed constructions are presented starting from a naive implementation that is used to illustrate the main ideas and the different constructions are analyzed theoretically. Finally, some potential optimizations for the proposed constructions are briefly discussed.
 
\subsection{Filters with a False Positive Free Set}

For a universe $U$ and a number $\FPP \in [0,1]$, a filter with a false positive free set is a randomised data structure constructed from two disjoint sets $S,T \subseteq U$. When queried for $x \in U$, it returns a bit $q(x) \in \{0,1\}$ such that
\begin{itemize}
    \item[(i)] $q(x) = 1$ for all $x \in S$,
    \item[(ii)] $q(x) = 0$ for all $x \in T$,
    \item[(iii)] $\Pr[q(x) = 1] \leq \FPP$ for all $x \in U \setminus (S \cup T)$, where “$\Pr$” refers to randomness used in the construction algorithm.
\end{itemize}
Note how this is a generalisation of several important problems. For $T = \emptyset$ we obtain an ordinary filter. For $U = S \cup T$ (or $\FPP = 0$) we obtain a set data structure. If we omit condition (iii) or $\FPP \geq 0.5$ then the problem has been referred to as the \emph{relative membership problem} \cite{CompressedStatic1} and calls for a static function with $1$-bit values.\footnote{Unless $|S| \approx |T|$, a \emph{compressed} static function should be used for space efficiency \cite{CompressedStatic1,StaticFunctions1,HKP:CompressedFunction:2009}.}
We do not claim any improvements for these widely studied special cases. From now on let $u = |U|$, $s = |S|$ and $t = |T|$. Our main interest concerns cases where $s$ and $t$ are both large, $u \gg s,t$ and $\FPP$ is bounded away from $0$ and $< 0.5$.

\subsection{Space Lower Bound}
Using arguments resembling those in \cite[Chapter 2.3]{M:Data_Structures:1984} for perfect hashing and in \cite{CarterBloombound} for Bloom filters, we now derive a space lower bound for filters with a false positive free set. Such a data structure implies a related data structure with essentially the same space consumption that satisfies (i), (ii) and the more combinatorial (and slightly weaker) property
\begin{equation}
    \tfrac{1}{u-s-t}|\{x \in U \setminus (S \cup T) \mid q(x) = 1\}| = \FPP' \tag{iii'}
\end{equation}
for some $\FPP' \leq \FPP$. The number $I$ of possible inputs $(S,T)$ for such a data structure is
\[ I = \binom{u}{s}\binom{u-s}{t} = \frac{u!}{s!\,t!\,(u-s-t)!}\]
Different inputs are not necessarily handled differently. Indeed, after construction, the data structure returns $1$ for a set $P$ of $s + \FPP'\cdot (u-s-t)$ elements of $U$ and $0$ for $U \setminus P$. As such, a single memory state is suitable for all inputs where $S \subseteq P$ and $T \subseteq U \setminus P$, i.e.\ for 
\[ W = \binom{s+\FPP'\cdot (u-s-t)}{s}\binom{(1-\FPP')\cdot (u-s-t)+t}{t}\]
inputs. By the pigeon hole principle, at least $I/W$ memory states are possible after construction and hence $\log_2(I/W)$ bits of memory are required. To obtain a simple bound, we assume that $u$ is large, more precisely $\min(\FPP',1-\FPP')\cdot u = \omega(\max(s^2,t^2))$. In that case
\begin{align*}
    \log_2(I/W) &= \log_2\big(\tfrac{u^{s+t}}{s!t!} / (\tfrac{(\FPP'u)^s}{s!}\cdot \tfrac{((1-\FPP')u)^t}{t!})\big)+ o(1)\\
    &= \log_2((\tfrac{1}{\FPP'})^s\cdot(\tfrac{1}{1-\FPP'})^t)+ o(1)\\
    &= s\cdot\log_2(\tfrac{1}{\FPP'})+ t\cdot\log_2(\tfrac{1}{1-\FPP'})+ o(1).\\
\end{align*}
When $\FPP'$ is small a good approximation is obtained using $\ln(\frac{1}{1-\varepsilon}) \geq \varepsilon$, namely
\[
    \log_2(I/W) \geq s\cdot\log_2(\tfrac{1}{\FPP'})+ t\cdot \FPP'\cdot\log_2(\mathrm{e})+ o(1).
\]
A lower bound for a filter with a false positive free set is now obtained by choosing $\FPP' \leq \FPP$ that minimises space consumption, i.e.
\begin{equation}
    \textsc{space} \geq \min\limits_{\FPP' \leq \FPP} s\cdot\log_2(\tfrac{1}{\FPP'})+ t\cdot \FPP'\cdot\log_2(\mathrm{e})+ o(1).\label{eq:lower-bound}
\end{equation}
Note that this suggests two regimes: When $t/s \ll \tfrac{\log_2(1/\FPP)}{\FPP}$ then using $\FPP' = \FPP$ is optimal and the term depending on $t$ is negligible. We recover essentially the lower bound $s\cdot\log_2(1/\FPP')$ for ordinary filters.
When $t/s$ is larger, however, the minimum is attained for a value $\FPP' < \FPP$. In particular the most compact filter with false positive free set should be expected to have a false positive rate $\FPP'$ \emph{smaller} than required. We will reencounter this phenomenon in our implementation which will, up to a constant factor, achieve the lower bound (\ref{eq:lower-bound}).









\subsection{High-level Design Strategies}

Before developing our preferred solutions, let us discuss a few high-level strategies that come to mind.
\begin{itemize}
    \item One might construct a filter for $S$ and---as an afterthought---\textbf{retouch} it \cite{RetouchedBF} such that no element from $T$ is a false positive, but without introducing false negatives. Applied to Bloom filters and xor filters, this plan seems hopeless.
    It is plausible, however, that it can work for adaptive filters. For example, in adaptive cuckoo or quotient filters \cite{ACF},\cite{Bender} by adapting the filter to the $T$ set before using the filter. The same applies to more recent variants of adaptive cuckoo filters \cite{KCP:SupportOptAdaptiveCuckoo:2021} where stored fingerprints that cause a false positive can be moved to alternative positions where they do not. However, removing one false positive has a chance of creating false positives for different elements in $T$, making it unclear how to guarantee the simultaneous removal of all false positives for $T$ efficiently. Even if this alternative works,
    our favoured approach is simpler and more space efficient.
    \item One could construct a \textbf{cascade of filters} as suggested for a similar setting in \cite{chazelle2004bloomier}. First, construct a filter for $S$. Then, construct a filter for those elements from $T$ that are false positives of the first filter. Then, construct a filter for those elements from $S$ that are false positives of the second filter (making them “false false positives”) and so on resulting in a sequence of ever smaller filters, each correcting the mistakes of the previous one. This would likely work, but our favoured approach is simpler and has better worst-case query time.
    \item One could use a $1$-bit static function with support $S \cup T$ and value $1$ for $x \in S$ and $0$ for $x \in T$. Unfortunately, most constructions return a uniformly random bit for $x \in U \setminus (S \cup T)$, giving us no easy way to control the false positive rate. An exception might be the construction from \cite{CompressedStatic1} where the output distribution for inputs $x \in U \setminus (S \cup T)$ coincides with the output distribution for random elements from the support. This yields $\FPP = \tfrac{s}{s+t}$ at first, but could plausibly be tuned for any false positive rate. We have not pursued this approach because it is rather complicated with no practical implementation that we know of.
    \item Bloomier filters \cite{chazelle2004bloomier} \emph{can} solve the problem at hand. Even though we improve upon the idea later in the paper, it is instructive to consider it in some detail in Subsection \ref{sec:NaiveApproach}.
\end{itemize}
Our favoured approach involves aspects of some of these ideas, combining a filter with a static function. Since xor-probing is suitable to implement both of these sub-tasks, it is the natural probing strategy for us to use.

\subsection{Naive construction}
\label{sec:NaiveApproach}

\begin{figure}[h]
  \centering
  \includegraphics[scale=0.55]{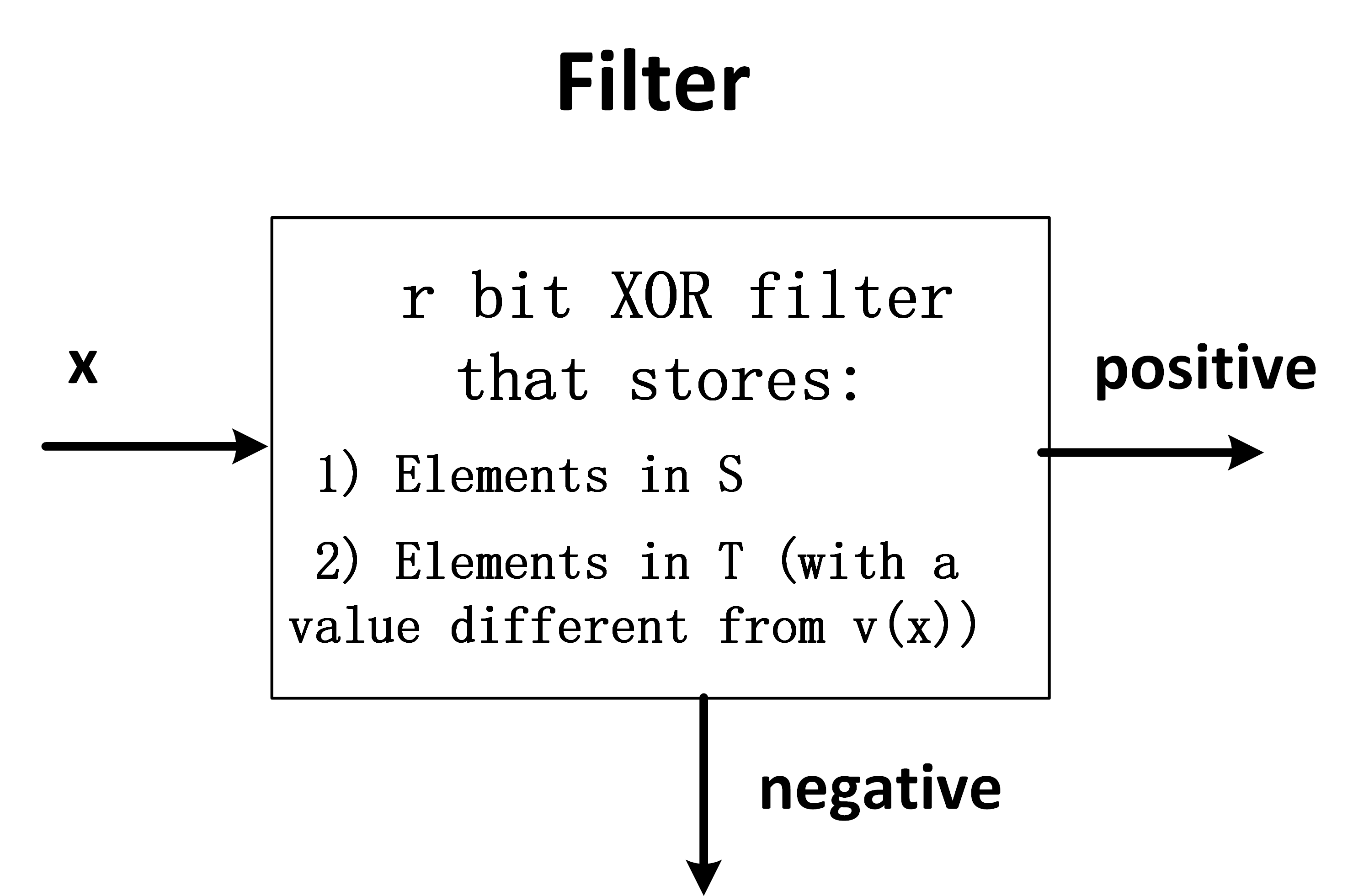}
  \caption{\small Diagram of the naive construction of a filter with a false positive free set $T$. 
  }\label{FigNaive}
\end{figure}

As discussed in the previous section, an xor probing filter ensures that for each element $x$ inserted, the filter computes a value $F_{xor}(x)$ that is equal to $v(x)$. Indeed a positive is returned when $F_{xor}(x) = v(x)$ and a negative otherwise. Therefore, to ensure that an element $x$ returns a negative, we can insert it on the filter with a value $v_n(x) \ne v(x)$. Then when searching for $x$ the filter will return $F_{xor}(x) = v_n(x)$ that is different from $v(x)$ so a negative is obtained.  This is illustrated in Figure \ref{FigNaive}. This naive construction can also be implemented with a Bloomier filter \cite{chazelle2004bloomier} using two categories---one for positives and one for negatives---and storing $S$ and $T$ on each of them respectively. 

This simple scheme can be used to insert all the elements in $T$ with values different from their $v(x)$ values and thus avoid false positives for $T$. However, this comes at a large cost in terms of memory as now the filter stores $|S \cup T|$ elements and thus requires approximately:  

\begin{equation}
    M_{naive} = (s+t)\cdot (1+\epsilon) \cdot r
\end{equation}

\noindent memory bits for a filter with a false positive rate of $2^{-r}$. This means that ensuring that each element on $T$ is not a false positive requires $(1+\epsilon) \cdot r$ bits, the same memory as for the elements in $S$. On the other hand, the check operation on the filter remains the same and thus removing the false positives has no implications on speed. 

Finally, let us consider the False Positive Probability (FPP) for elements of the universe $U$ that are not in $S \cup T$. For those, the value returned by the filter $F_{xor}(x)$ would look like a uniformly distributed random value and thus the probability that it matches the value of the element $v(x)$ and thus a false positive is obtained will be approximately $2^{-r}$.

\subsection{Two filters (TF) construction}
\label{TwoFilter}

The memory footprint can be significantly reduced by observing that to ensure that an element is not a false positive, using a one bit filter suffices. Therefore, an $r-1$ bit filter can be used to store the elements in $S$ and a second one bit filter to store the elements in $S \cup T$ (the elements in $T$ with their negated bit value). In fact, this second one-bit filter can be seen as a static function that maps elements in $S$ to 1 and elements in $T$ to 0. Based on a subtle observation, not all elements in $T$ have to be inserted on the second filter. Indeed only the elements in $T$ that return a false positive on the first filter need to be inserted on the second filter. Let us denote as $F$ the set formed by the elements in $T$ that return a false positive in the first filter, and let $f = |F|$. Then only the elements in $F$ have to be inserted on the second filter and $f \approx \frac{t}{2^{r-1}}$, so $F$ is typically much smaller than $T$. Therefore, this scheme would require: 

\begin{equation}
   \label{MTinitial}
    M_{TF_{initial}} = s \cdot (1+\epsilon) \cdot (r) + f \cdot (1+\epsilon) 
\end{equation}

\noindent memory bits and can be expressed as a function of $T$ as: 

\begin{equation}
     M_{TF_{initial}}  = s \cdot (1+\epsilon) \cdot (r) + \frac{t}{2^{r-1}} \cdot (1+\epsilon) 
\end{equation}

\noindent which is much lower than that of the naive construction (using assumption $r > 1$ from assumption $\FPP < 0.5$).

\begin{figure}[h]
  \centering
  \includegraphics[scale=0.55]{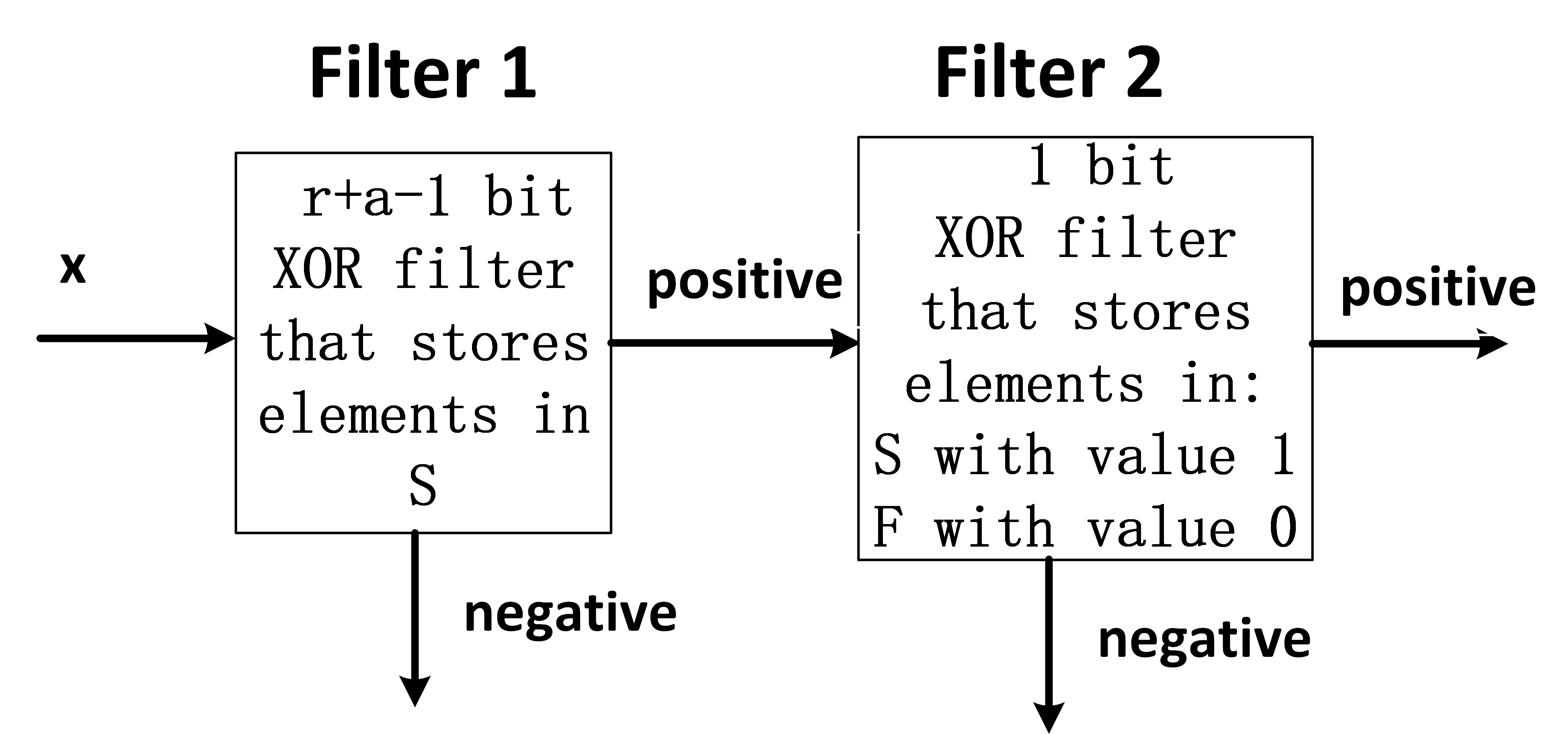}
  \caption{\small Diagram of the Two Filter (TF)  construction with a false positive free set $T$. 
  }\label{FigTF}
\end{figure}

Looking carefully at the second term of equation \ref{MTinitial}, it can be observed that each element that needs to be removed requires $(1+\epsilon)$ bits. This is not efficient when $\frac{t}{2^{r-1}} > 2 \cdot s$. To explain why, for example, let us consider that  $t= 3 \cdot 2^{r-1}\cdot s$ so that $f = 3 \cdot s$, then we would need $(1+\epsilon) \cdot (r+3) \cdot s$ memory bits. Instead, let us consider an alternative configuration in which the first filter has $r$ bits instead of $r-1$ bits. Then, we would have $f = \frac{3}{2} \cdot s$ and only $(1+\epsilon) \cdot (r+2.5) \cdot s$ memory bits are needed. This shows that when $\frac{t}{2^{r-1}} > 2 \cdot s$, it is more efficient to add bits to the first filter using $r' = r+a$ such that $\frac{t}{2^{r+a-1}} < 2 \cdot s$. This is so because by adding an additional bit to the first filter we reduce in half the size of $F$ and thus when $F$ is large this is more efficient than storing all the elements in $F$ in the second filter. Therefore, in the TF construction when that occurs, $a = \lceil \log_2(\frac{t}{2^{r-1} \cdot s})\rceil -1$ bits are added to the first filter to reduce the size of $F$. This generalized TF construction is illustrated in Figure \ref{FigTF}. 

The memory requirements are now: 

\begin{equation}
    M_{TF} = s \cdot (1+\epsilon) \cdot (r+a) + f \cdot (1+\epsilon) 
\end{equation}

\noindent that can be expressed as a function of $T$ as: 

\begin{equation}
    M_{TF} = s \cdot (1+\epsilon) \cdot (r+a) + \frac{t}{2^{r+a-1}} \cdot (1+\epsilon) 
\end{equation}

\noindent so much smaller than in the naive construction. The FPP would be approximately $2^{-(r+a)}$ so the use of additional bits on the first filter has the side benefit of reducing the FPP.

A drawback of the TF construction is the speed overhead of checking two filters for some queries, specifically positive and a fraction of negative queries. This would be acceptable in many cases as negative queries motivate the choice to use a filter. 

\subsection{Integrated filters (IF) construction}
\label{IntegratedFilter}

\begin{figure}[h]
  \centering
  \includegraphics[scale=0.55]{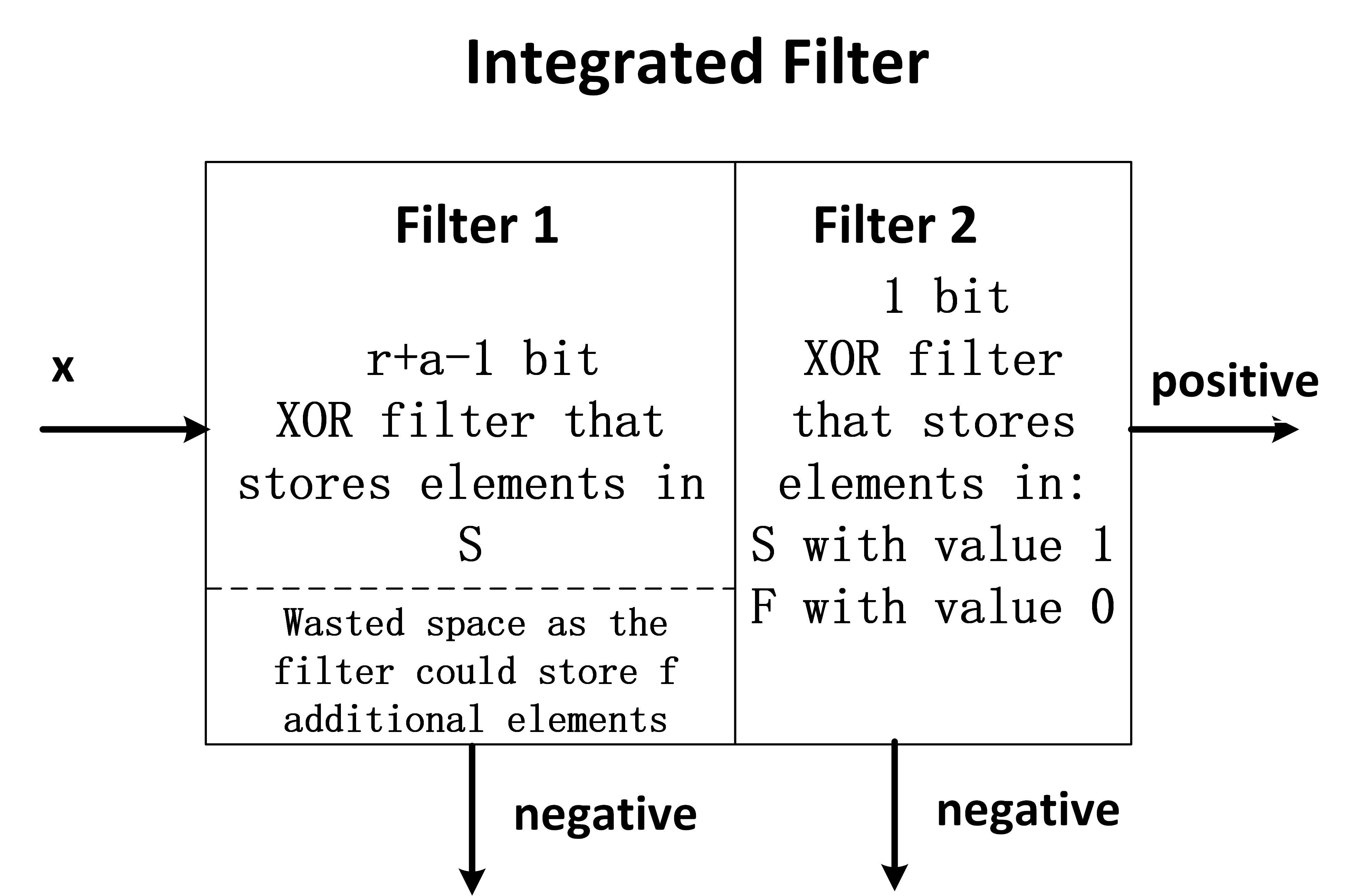}
  \caption{\small Diagram of the Integrated Filter (IF) construction of a filter with a false positive free set $T$ when $c= 1$. Filters 1 and 2 use the same hash functions to compute the addresses to access and have the same size. 
  }\label{FigIF1}
\end{figure}

The two filters construction is effective to reduce the memory footprint (space) but introduces other overheads (time) that may be an issue in some applications. Therefore, it would be beneficial to have a construction that is as fast as a single filter and is more memory efficient than the naive construction requiring $(s + t) \cdot (1 + \epsilon)$ positions. By increasing the number of positions in filter 1 to match filter 2, from $s \cdot (1 + \epsilon)$ to $(s + f) \cdot (1 + \epsilon)$ positions, both filters can be fully integrated in the same data structure. And by using the same position hash functions for both\footnote{This leads to potential correlation in construction success probability, not in false positive probability.}, the bits for both filters can be concatenated on a single larger entry per position, so that querying both involves no more memory lookups than a single filter query. This scheme is illustrated in Figure \ref{FigIF1} and would require a memory footprint of:

\begin{equation}
    \label{MIF}
    M_{IF}|_{c=1} = (s+\frac{t}{2^{r+a-1}}) \cdot (1+\epsilon) \cdot (r+a)  
\end{equation}

\noindent In this case, adding bits to the first filter is more beneficial as the cost of the elements in $F$ is larger. The analytical derivation of the optimal value for $a$ becomes more complex as it now appears twice on the equation but it can be easily found by computing equation \ref{MIF} starting with $a= 0$ and increasing $a$ until the memory required increases. Alternatively, when $r$ is much larger that $a$, we can approximate $r+a$ with $r$ and then the condition to minimize the memory becomes becomes $\frac{t \cdot (r-1)}{2^{r+a-1}} > 2 \cdot s$ and thus: 

\begin{equation}
    a = \lceil \log_2(\frac{t \cdot (r-1)}{2^{r-1} \cdot s})\rceil -1
\end{equation}

This results in a memory overhead that is larger than that of the TF approach, especially when the initial $F$ is large.

\begin{figure}[h]
  \centering
  \includegraphics[scale=0.55]{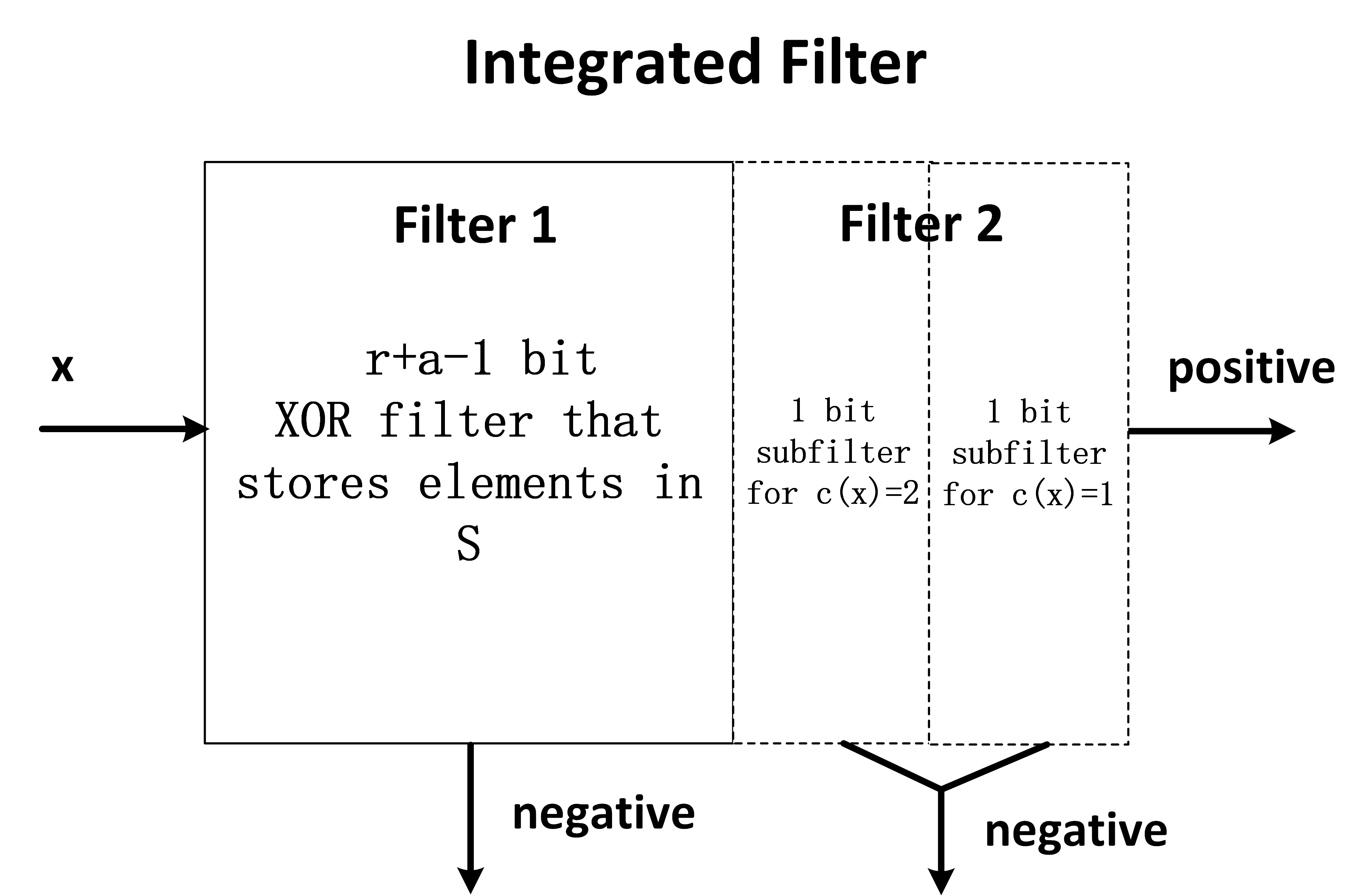}
  \caption{\small Diagram of the Integrated Filter (IF) construction of a filter with a false positive free set $T$ when $c=2$. Filter 2 is composed of two subfilters each element $x$ maps to one of them given by $c(x)$.  
  }\label{FigIF2}
\end{figure}

To reduce the overhead, the previous construction can be generalized as shown in Figure \ref{FigIF2}. It can be seen that now $c$ bits are used for the second filter that is decomposed in $c$ subfilters each storing a subset of the elements in $S \cup F$. As in the initial construction, the bits are concatenated per position, each having $r+a-1$ bits for the first filter and $c$ for the second. In more detail, another hash function is used to compute $c(x)$ that selects the subfilter in the second filter. Therefore each subfilter stores approximately $\frac{|S \cup F|}{c}$ elements. When this is smaller than $s$, the memory footprint is:

\begin{equation}
    M_{IF}|_{c>1} = s \cdot (1+\epsilon) \cdot (r+c-1).  
\end{equation}

The minimum value of $c$ that can be used to ensure that $\frac{|S \cup F|}{c} < s$ given a set $T$ can be computed as:

\begin{equation}
  \label{eqfmin}
  c_{min} = 1+ \big\lceil \frac{t}{s \cdot 2^{r-1}} \big\rceil  
\end{equation}

So the minimum memory needed is given by:

\begin{equation}
    M_{IF}|_{c_{min}} = |S| \cdot (1+\epsilon) \cdot (r + \big\lceil \frac{t}{|S| \cdot 2^{r-1}} \big\rceil )  
\end{equation}

As in the previous constructions, a subtle observation is that when the value obtained for $c_{min}$ is larger than two, it is better to increase the value of $r$ and use $c=2$. This is better seen with an example, let us assume that $\frac{t}{s \cdot 2^{r-1}}$ is equal to 3.6. Then $c_{min}$ would take a value of five, so five bits have to be added the filter. Instead if we use $r'=r+2$, then $\frac{t}{s \cdot 2^{r'-1}}$ would be 0.9 and $c_{min}$ would take a value of two, needing in total four bits, so fewer than the original five. In general, by adding $a$ bits to $r$ to obtain $r'$, the following number of bits would be needed:

\begin{equation}
  \label{eqra}
   a + 1 + \big\lceil \frac{t}{s \cdot 2^{r+a-1}} \big\rceil  
\end{equation}

when this value is lower than that of Equation \ref{eqfmin}, adding bits to $r$ is more efficient. 

Therefore, the procedure to select the parameters for the IF design is as follows:

\begin{enumerate}
    \item Determine $c_{min}$ according to equation \ref{eqfmin}.
    \item If $c_{min} = 2$ no additional bits are needed for $r$, the filter is implemented with the original value of $c_{min} = 2$.
    \item If $c_{min} > 2$ add $a= \lceil \log_2(c_{min}-1) \rceil$  bits to $r$ and recompute equation \ref{eqfmin} to obtain the final value of $c_{min} = 2$.
    \item Implement the IF with $r'= r+a$ bits on the first filter and $c=2$ subfilters in the second filter.
\end{enumerate}

So, it can be seen that in all cases $c=2$ is used. When $a=1$, for example when $\frac{t}{s \cdot 2^{r-1}}$ is equal to 1.8, both alternatives require the same number of bits (four), but adding the bit to $r$ has the side benefit of reducing the false positive rate for elements that are not in $T$ so it should be used. 

In summary, in the integrated filter construction to reduce the memory footprint in a given design, the value of $c$ is always fixed to two and $a$ is computed. In terms of lookups, this construction does not introduce additional memory accesses, only some operations to select and check the subfilter when needed. 

The FPP would also be approximately $2^{-(r+a)}$ so similar to previous constructions when $a=0$ and lower otherwise. 


\subsection{Analysis}

\begin{figure}[h]
  \centering
  \includegraphics[scale=0.65]{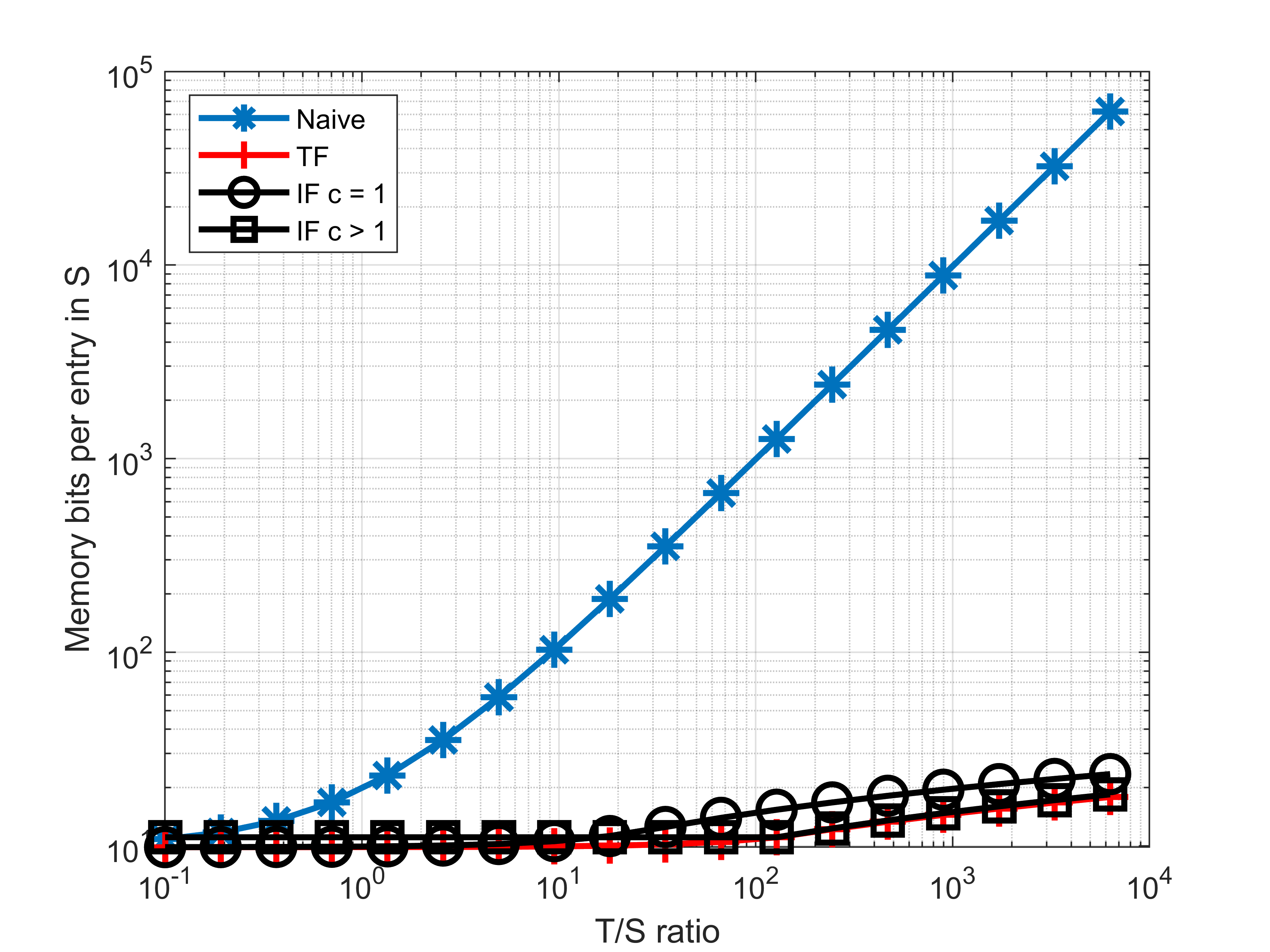}
  \caption{\small Memory bits per element in $S$ to implement the different FPFS filters for different sizes of $T$ when $r=8$ and for $S$ of size one million elements.  
  }\label{FigMem81}
\end{figure}

Let us now compare the different filters with a false positive free set (FPFS) constructions presented in the previous subsections. We consider a set $S$ of size one million elements, $\epsilon=0.23$, $r=8$ and different sizes of the FPFS set $T$ and compute the memory required by the three constructions using the equations derived in the previous subsections. The results are shown in Figure \ref{FigMem81}. It can be seen that the naive construction requires much more memory than the others as the size of $T$ increases confirming that it is not an efficient construction unless $T$ is very small. The same reasoning applies to a Bloomier based construction and therefore none of the two is considered further in the rest of the paper.  

\begin{figure}[h]
  \centering
  \includegraphics[scale=0.65]{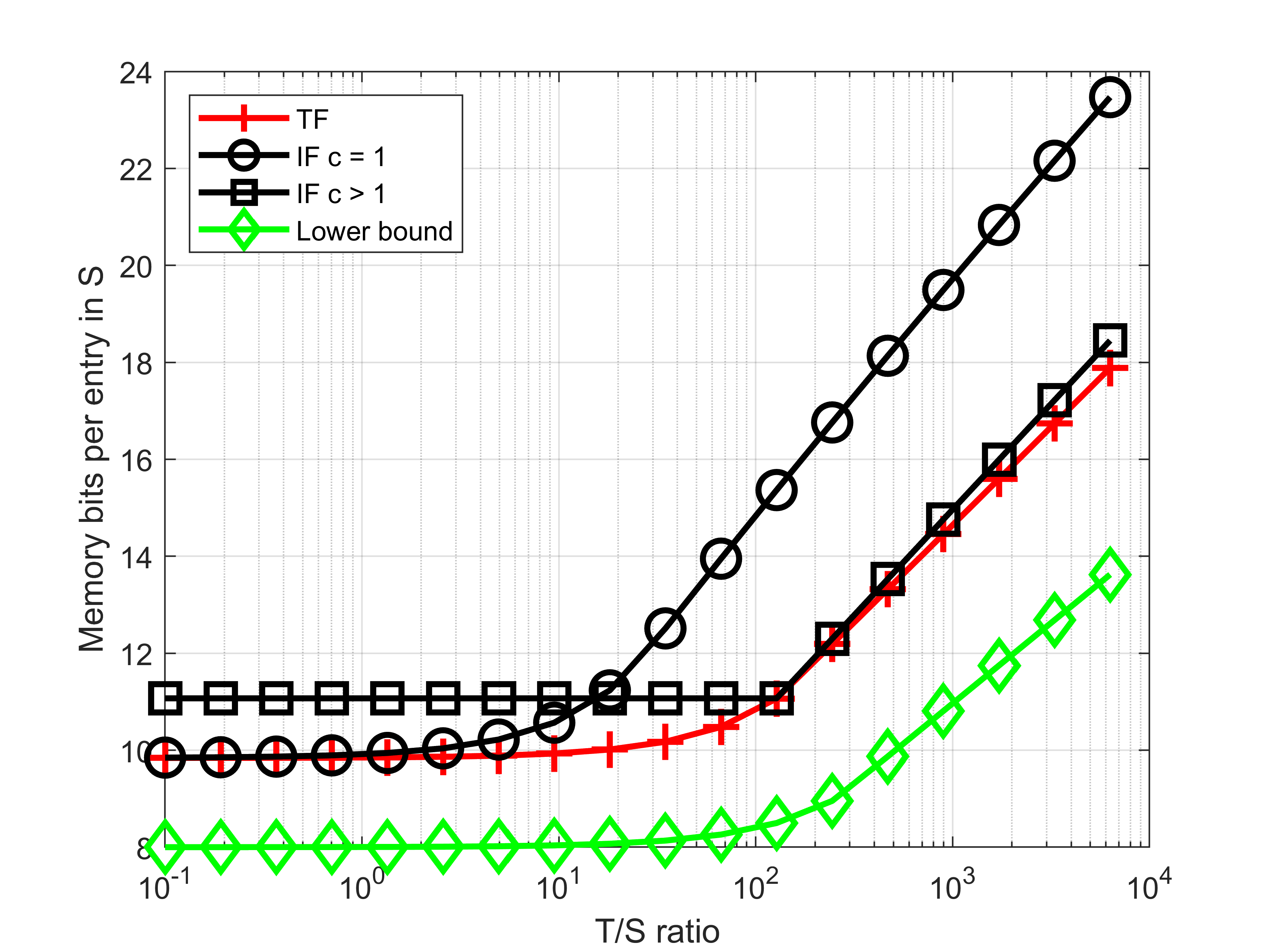}
  \caption{\small Memory bits per element in $S$ required to implement the TF and IF FPFS filters for different sizes of $T$ when $r=8$ and for $S$ of size one million elements.  
  }\label{FigMem82}
\end{figure}

\begin{figure}[h]
  \centering
  \includegraphics[scale=0.65]{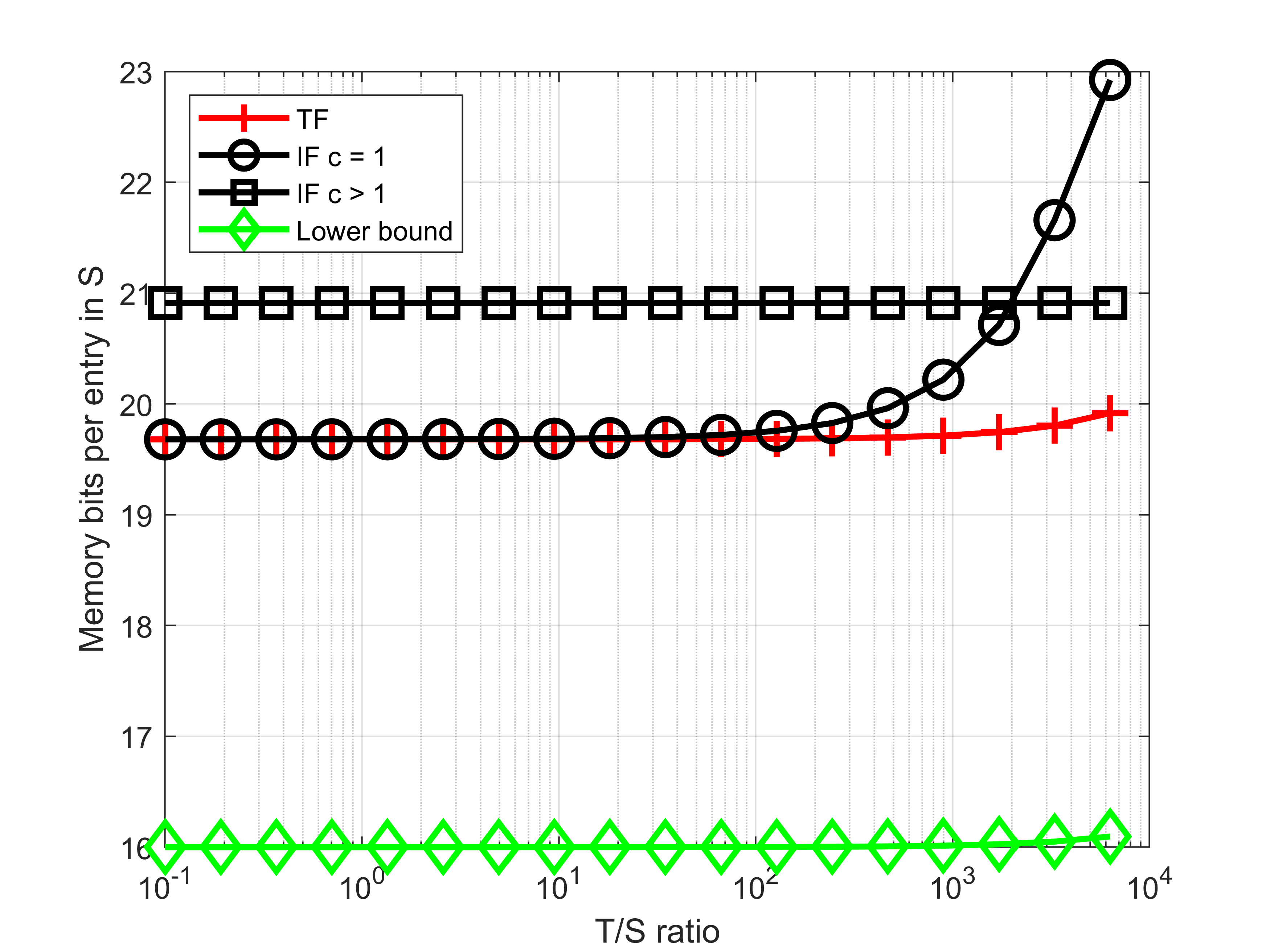}
  \caption{\small Memory bits per element in $S$ required to implement the TF and IF FPFS filters for different sizes of $T$ when $r=16$ and for $S$ of size one million elements. 
  }\label{FigMem162}
\end{figure}

To better compare the other two constructions, their results are shown in Figure \ref{FigMem82} for $r=8$ and in Figure \ref{FigMem162} for $r=16$ that also show lower bound (\ref{eq:lower-bound}). When $r=8$, TF is the most efficient construction in terms of memory while IF with $c=1$ is more effective when $T$ is small while the IF with $c>1$ is better when $T$ is large. As the size of $T$ increases, the memory required also increases and more bits are added to the first filter. When $r=16$, the use of memory is similar in most cases (except when $T$ is very large for the IF with $c=1$) as the size of $F$ is very small (recall that $c= \frac{t}{2^{r+a-1}}$). Finally it can be seen that the proposed filters are reasonably close to the lower bound being the difference mostly due to the $\epsilon$ =23\% memory overhead introduced by xor filters \cite{XORfilter}.

An interesting observation is that the IF construction offers an efficient alternative for all values of $T$ by selecting the appropriate value of $c$. Therefore, the integrated construction can be used to implement FPFS filter with a small memory footprint and lookups that require the same number of memory accesses as the original xor probing filter from which they are derived.

\subsection{Potential Optimizations}

In this section we discuss two potential optimizations for the proposed filter constructions. The evaluation and further analysis of these optimizations are left as future work and not considered further in this paper.

\subsubsection{Hybrid of TF and IF}
We observe that a hybrid of the TF and IF approaches could achieve the memory footprint of TF with the speed of IF for a portion of queries. This could be useful in speeding up positive queries when TF space efficiency is desired. Essentially, the hybrid would not round up in computing $c_{min}$ (Equation~\ref{eqfmin}), would separate off and size the $c(x) = 2$ subfilter for only $f$ elements, and would make $c(x)$ a weighted hash, such that $c(x) = 1$ for approximately $s$ random elements of $S \cup F$ and $c(x) = 2$ for the approximately $f$ remaining elements. 

\subsubsection{Generalized IF}
We observe that when minimal rounding is involved in choosing $c_{min}$ (Equation~\ref{eqfmin}), IF is as memory efficient as TF, as in Figure \ref{FigMem82} near $T/S = 10^{2}$. With rounding, only $s+f$ elements are stored where there is space for $2s$ elements. If an independently hashed $\frac{s - f}{s}$ subset of elements map to both subfilters, this space can be filled for the purpose of lowering the false positive rate of elements not in $S \cup T$. Note that to guarantee false positive freedom, elements in $F$ only need to be added to one subfilter even if mapping to both. The approach of mixing the matching of $r$ and $r+1$ bits is known to incur only small space efficiency overheads compared to an optimal approach with truly fractional $r$ \cite{Ribbonfilter}. However, TF and generalized IF do not compare so directly because they offer different choices for false positive rates, except when the structure is standard IF with the same memory efficiency as TF because $t$ is a power of two. 

\section{Evaluation}
\label{Evaluation}

The proposed filters with a FPFS from Sections \ref{TwoFilter} and \ref{IntegratedFilter} have been implemented using as base filters the Java implementation of the xor filter \cite{XORfilter}\footnote{The code is available in https://github.com/amacian/fastfilterfpfs}. Then several experiments have been conducted to validate the filters and measure their performance. All simulations have been run on a Windows 10 machine with an Intel(R) Core(TM) i7-10700 CPU running at 2.90GHz. 


The first experiment focuses on validating the filter functionality by testing that all elements in $S$ return a positive, all elements in $T$ a negative and the rest have a false positive probability that is approximately $2^{-(r+a)}$. This has been done for a set $S$ of one hundred thousand elements and different sizes of $T= 10K,100K,1M$ and $10M$ elements when $r=4,8,16$. First, sets $S$ and $T$ were generated using random elements. Then, the filters were constructed and then all elements in $T$ were queried, in all cases there were no false positives. Therefore, the proposed filters provide the false positive free set feature. The memory usage and false positive rate for elements not in $T$ were measured.

\begin{figure*}[h]
  \centering
  \includegraphics[scale=0.36]{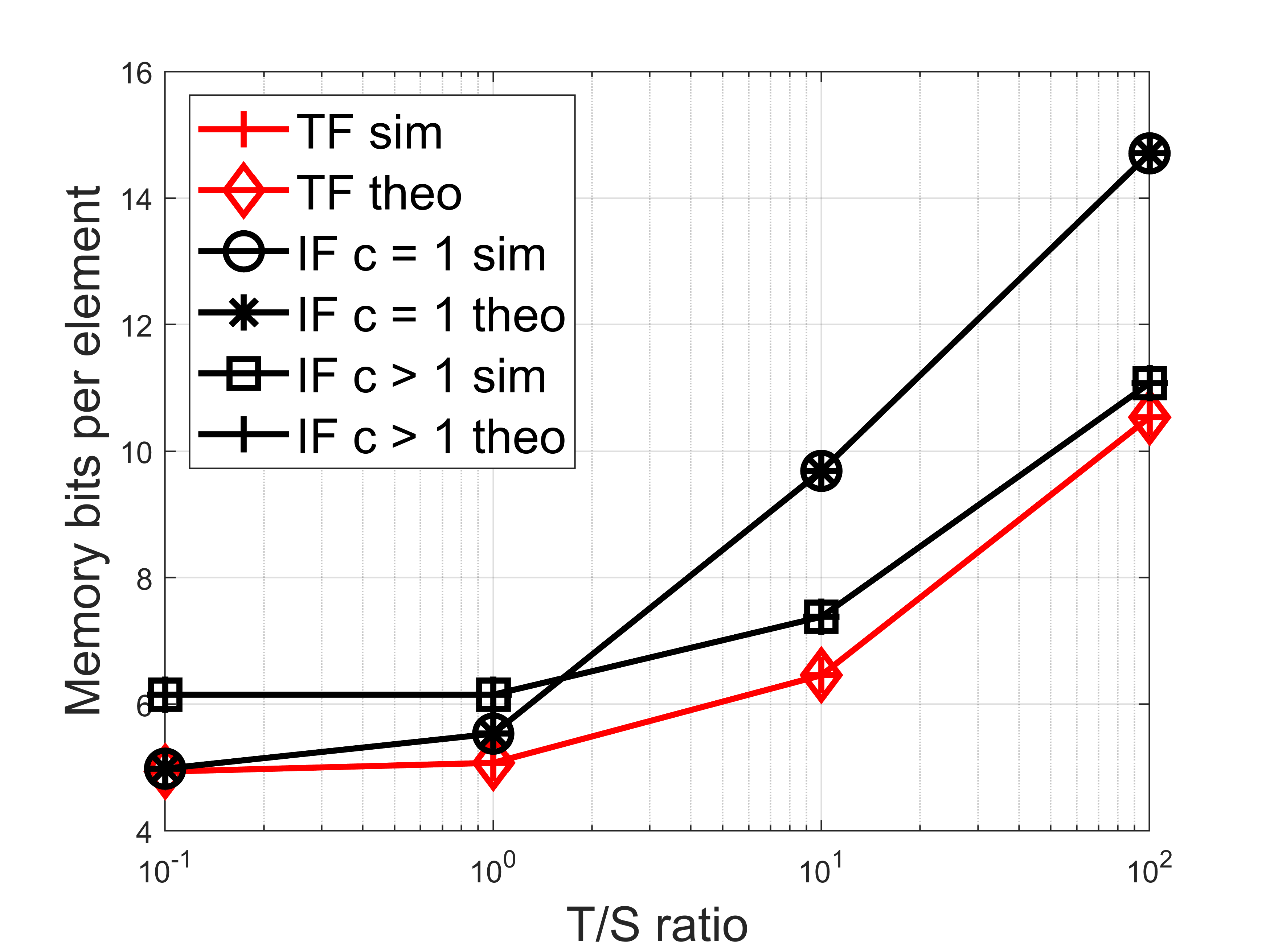}
  \includegraphics[scale=0.36]{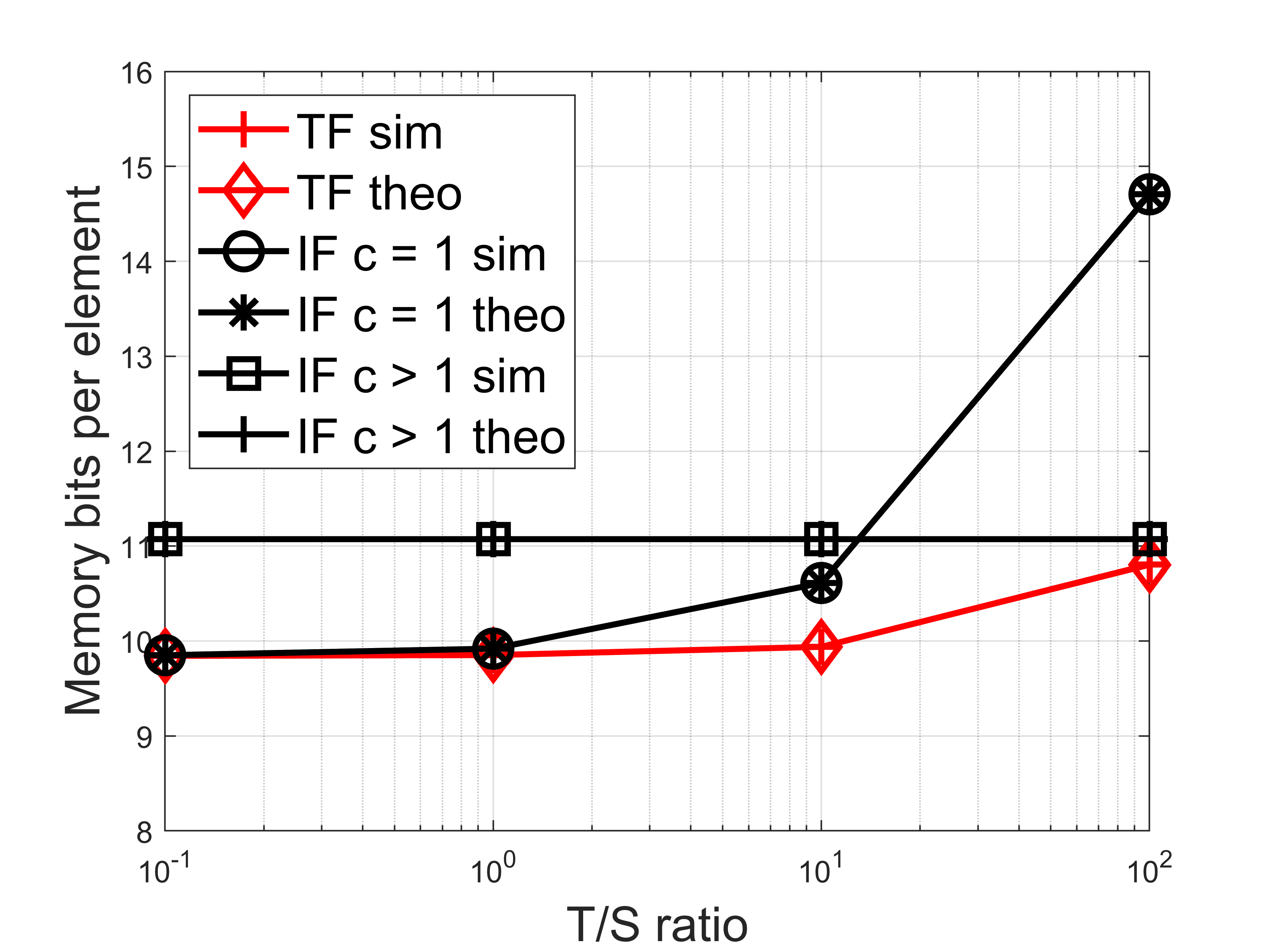}
  \includegraphics[scale=0.36]{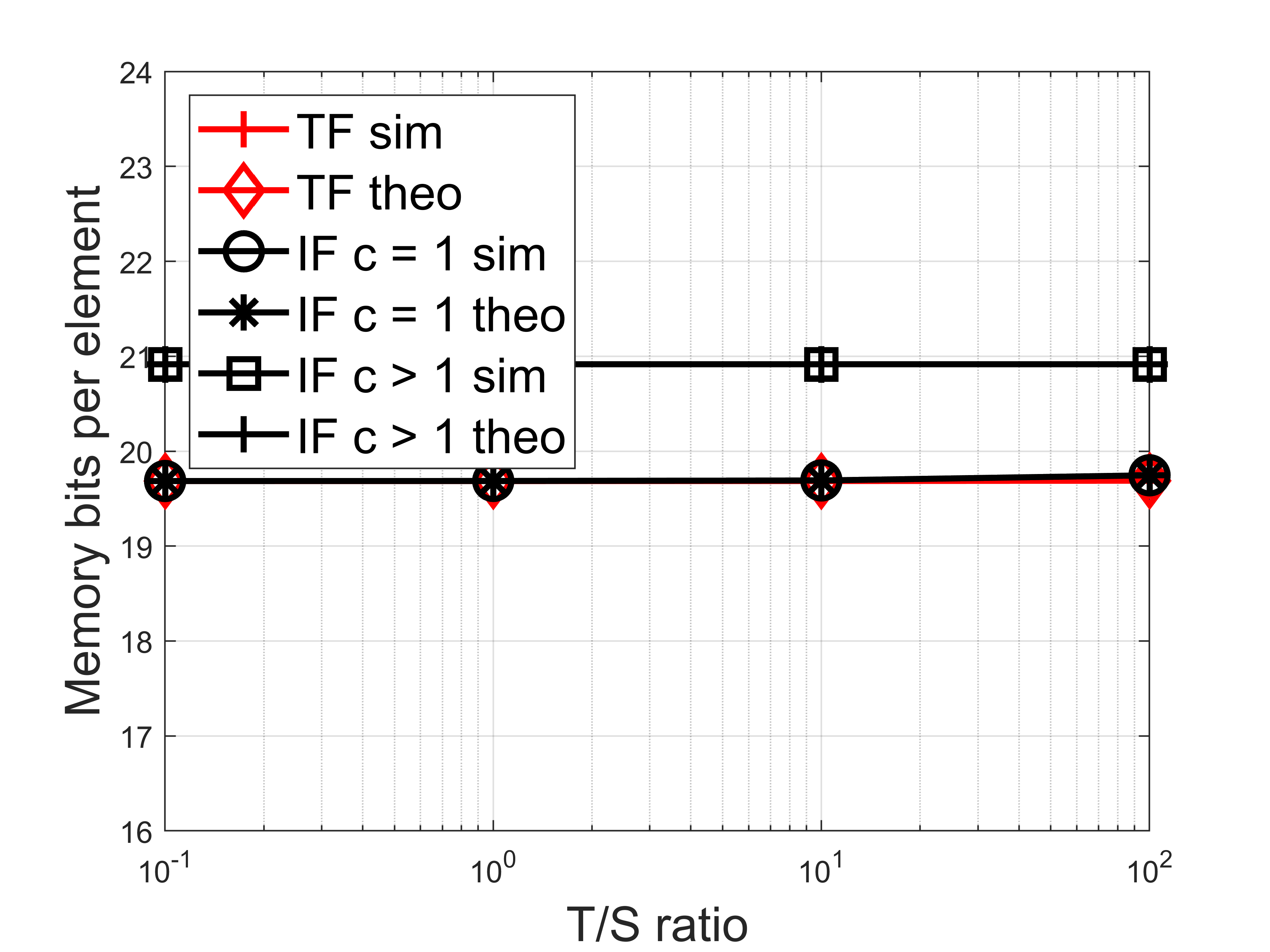}
  \caption{\small Memory required to implement the TF and IF FPFS filters for different sizes of $T$ when $r=4$ (left), $r=8$ (middle), $r=16$ (right) and for $S$ of size one hundred thousand elements.  
  }\label{FigExp1Mem}
\end{figure*}

The results for memory usage are shown in Figure \ref{FigExp1Mem} that includes the theoretical estimates presented in the previous section. The values for an Xor filter are similar to those of the TF scheme when $T$ is small that correspond to the leftmost points on the plots. It can be seen that the simulation results match the theoretical analysis. The TF construction is the most efficient in terms of memory usage as expected but the IF construction can also be implemented with a reasonable amount of memory by appropriately selecting the value of $c$. The results for the naive construction are not shown as much more memory is required making the filters not practical when the number of elements if $T$ is large.   

\begin{figure*}[h]
  \centering
  \includegraphics[scale=0.36]{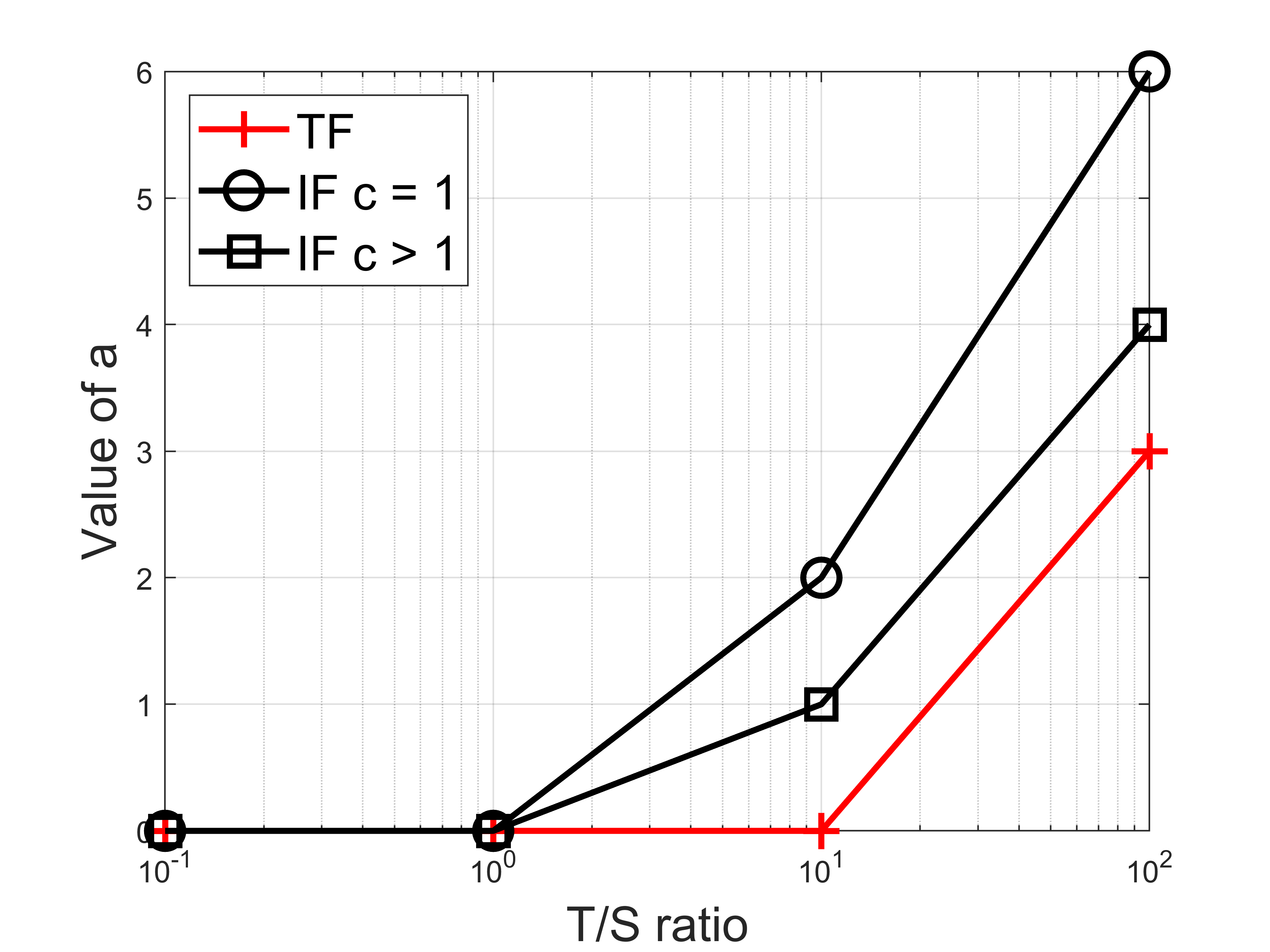}
  \includegraphics[scale=0.36]{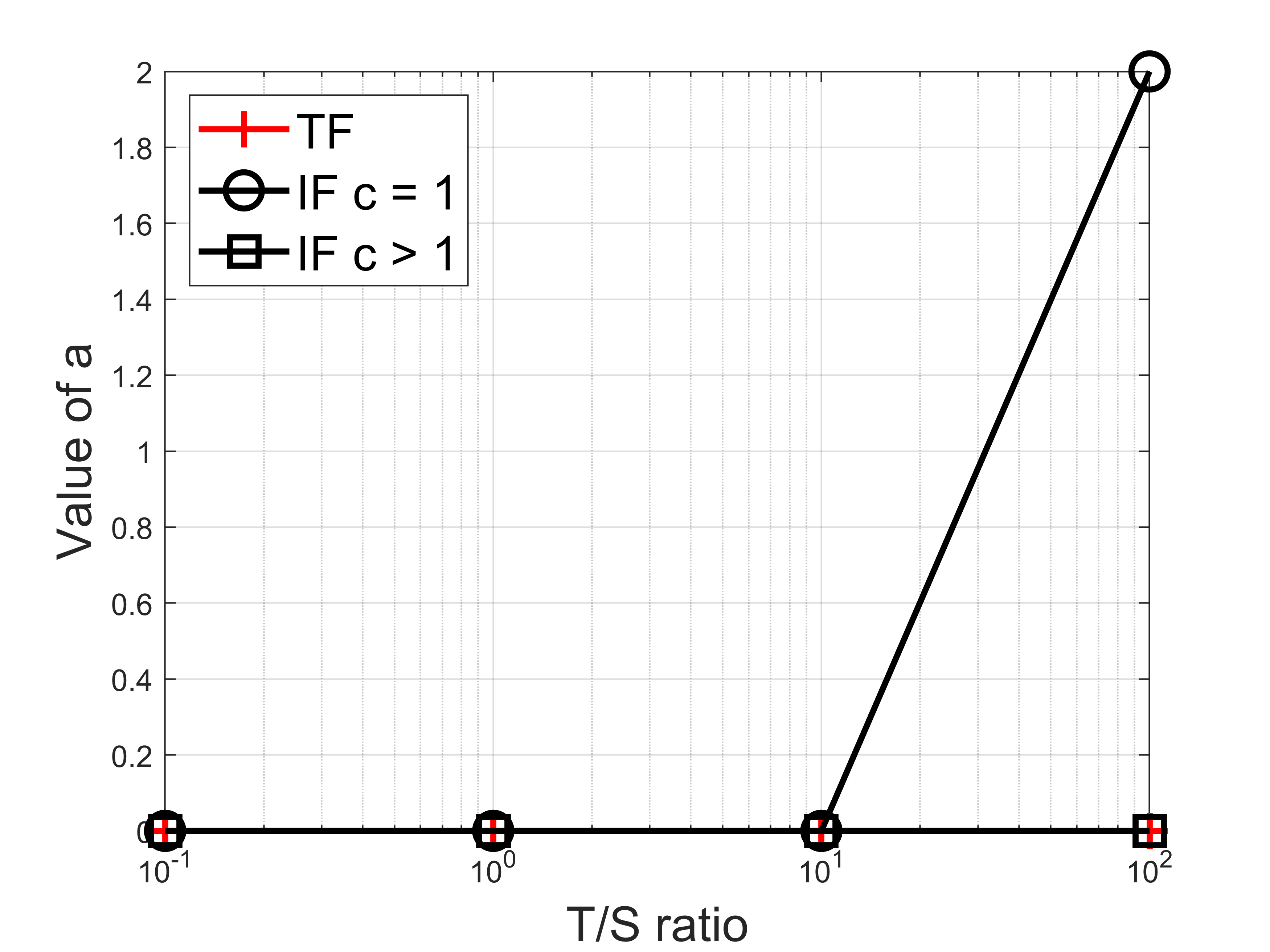}
  \includegraphics[scale=0.36]{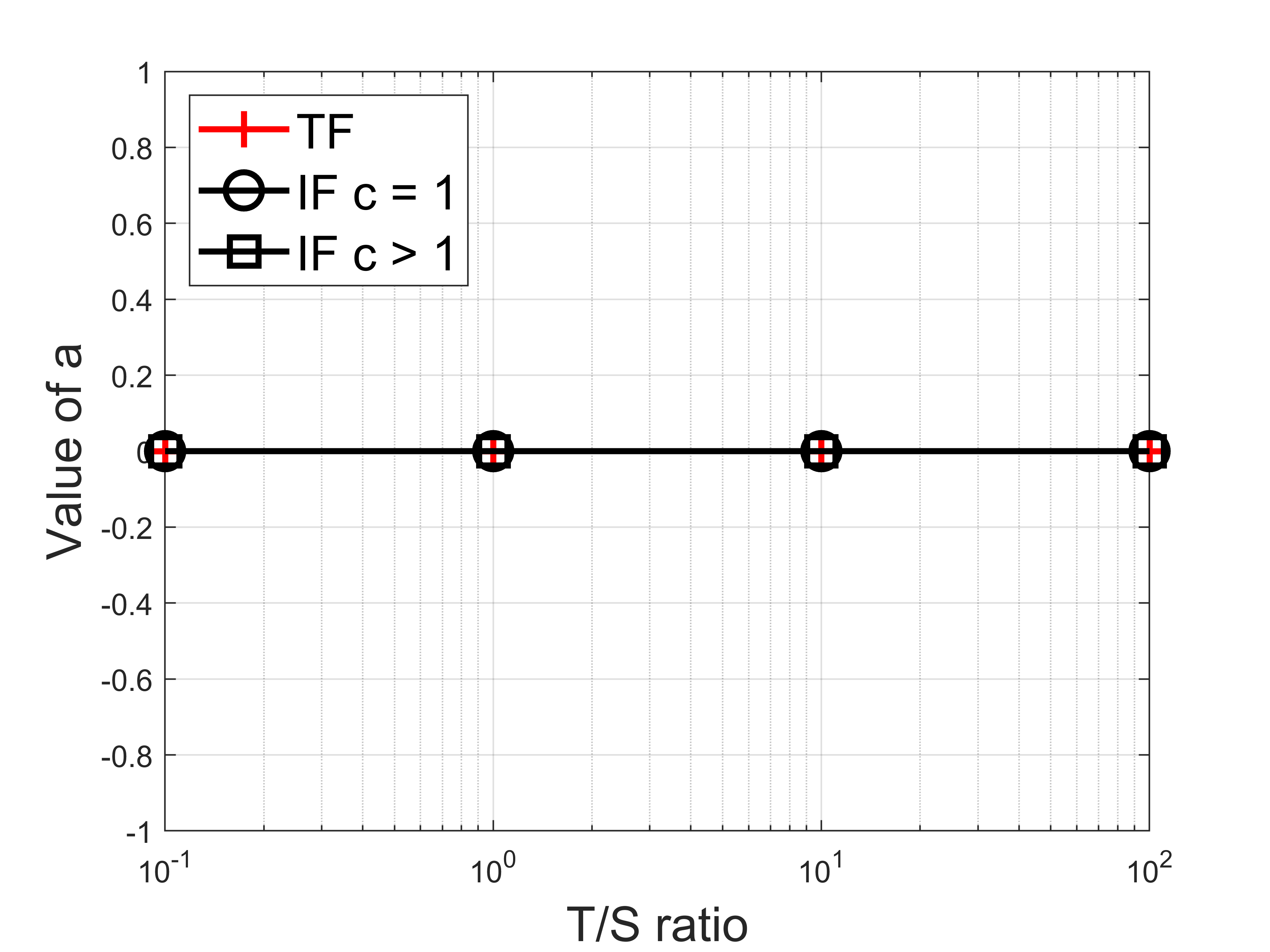}
  \caption{\small Value of $a$ for the TF and IF FPFS filters for different sizes of $T$ when $r=4$ (left), $r=8$ (middle), $r=16$ (right) and for $S$ of size one hundred thousand elements.  
  }\label{FigExp1a}
\end{figure*}

The parameter $a$ selected for each configuration was also logged and the results are shown in Figure \ref{FigExp1a}. It can be seen that when $r=16$, the value of $a$ is always zero as the size of $F$ is much smaller than that of $S$. Instead, when $r=4,8$ values larger than zero are selected as the size of $T$ grows. 

\begin{figure*}[h]
  \centering
  \includegraphics[scale=0.36]{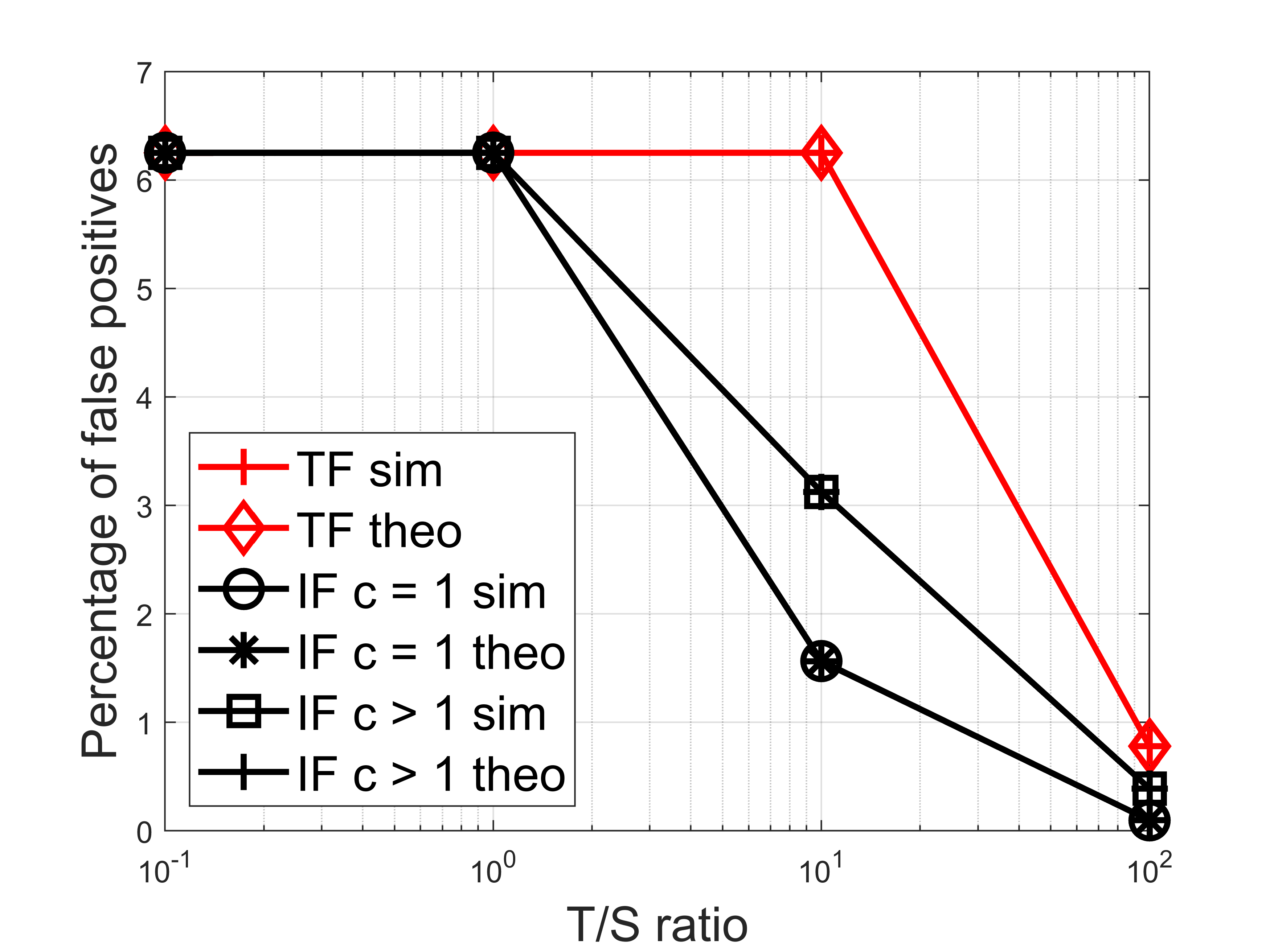}
  \includegraphics[scale=0.36]{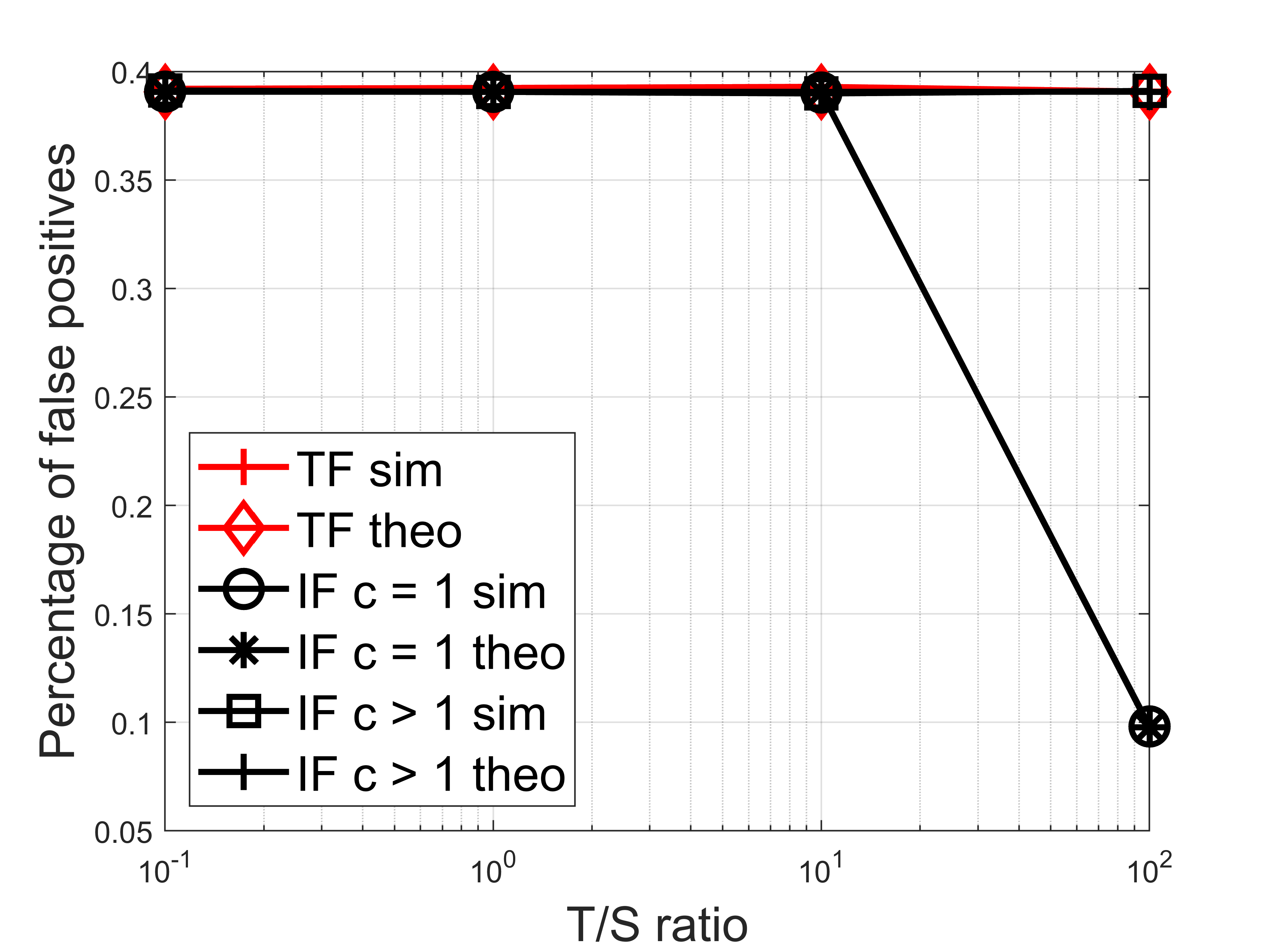}
  \includegraphics[scale=0.36]{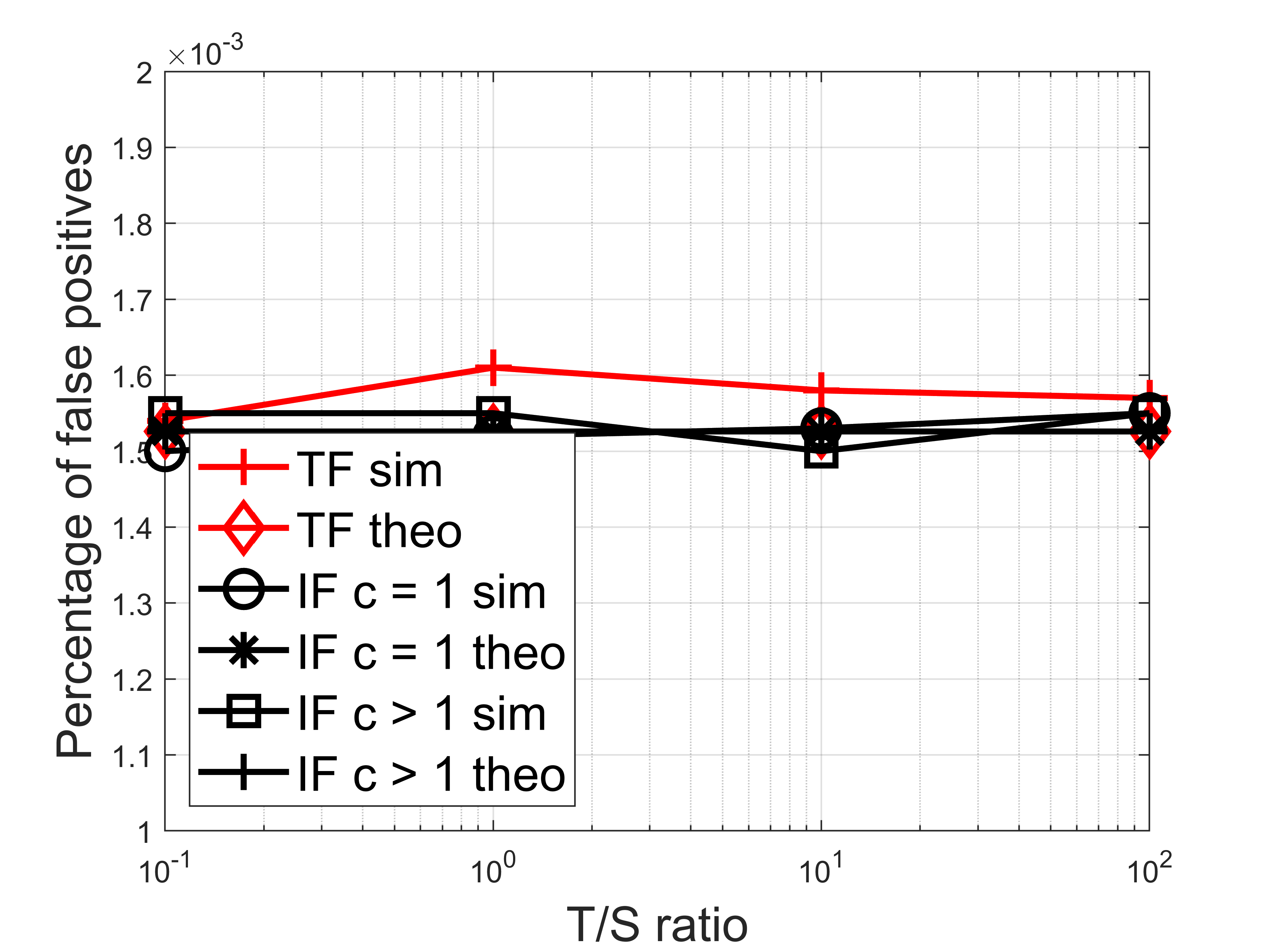}
  \caption{\small Fraction of false positives for negative elements not in $T$ for the TF and IF FPFS filters for different sizes of $T$ when  $r=4$ (left), $r=8$ (middle), $r=16$ (right) and for $S$ of size one hundred thousand elements.  
  }\label{FigExp1FPP}
\end{figure*}

Finally, the false positive probability for negative elements that are not in $T$ was measured with ten million random elements. The results are summarized in Figure \ref{FigExp1FPP} that also shows the theoretical estimate given by $\frac{1}{2^{r+a}}$. It can be seen that simulation results match the theoretical analysis.

The second experiment evaluates the speed of the proposed FPFS filter. As the absolute performance values are dependent on the platform used to run the experiments and also on the programming language and tools, we also include the results for the base Xor filter Java implementation \cite{XORfilter}. This enables a relative comparison in terms of the overhead introduced by our FPFS filters. 

\begin{figure*}[h]
  \centering
  \includegraphics[scale=0.36]{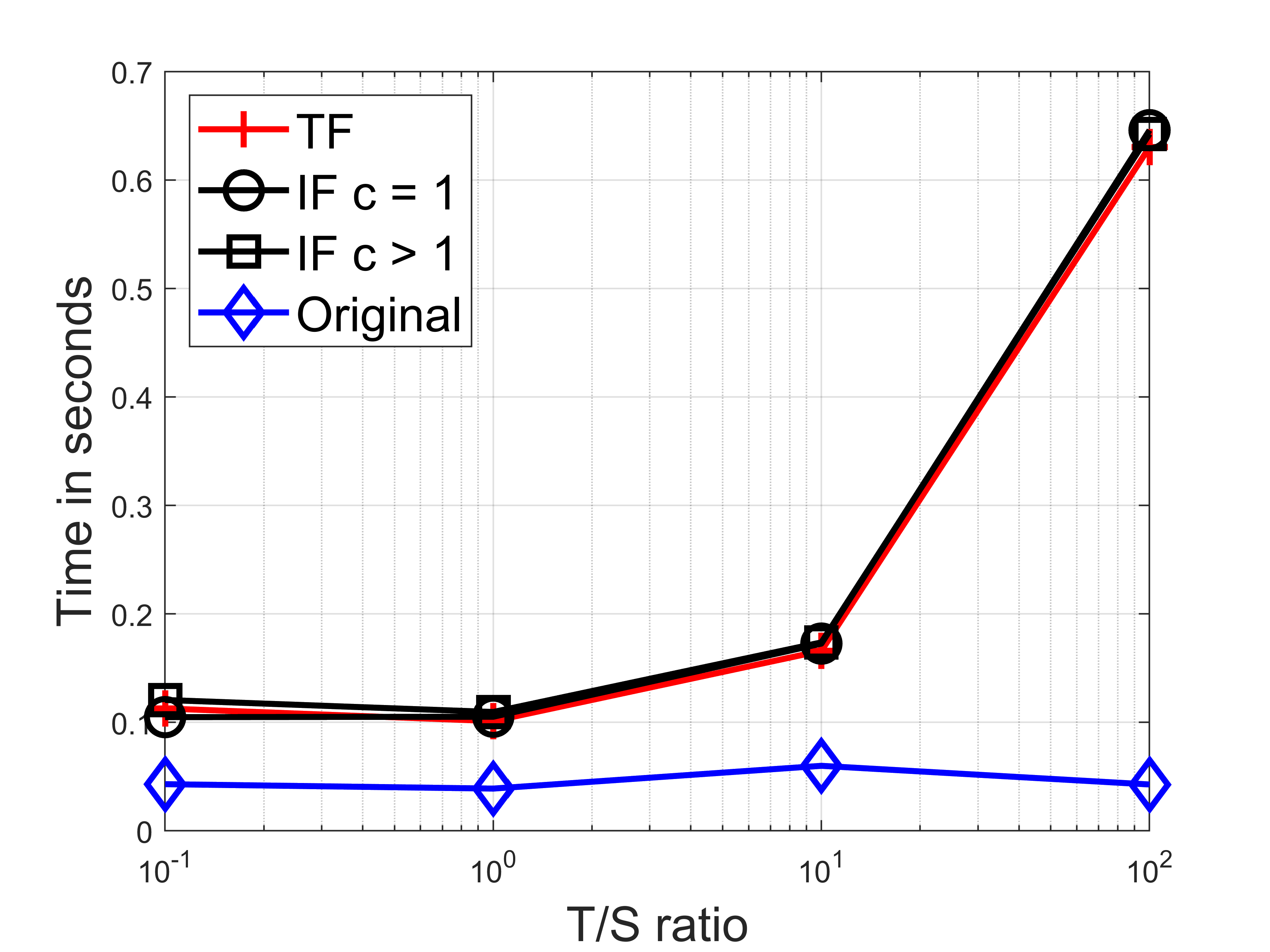}
  \includegraphics[scale=0.36]{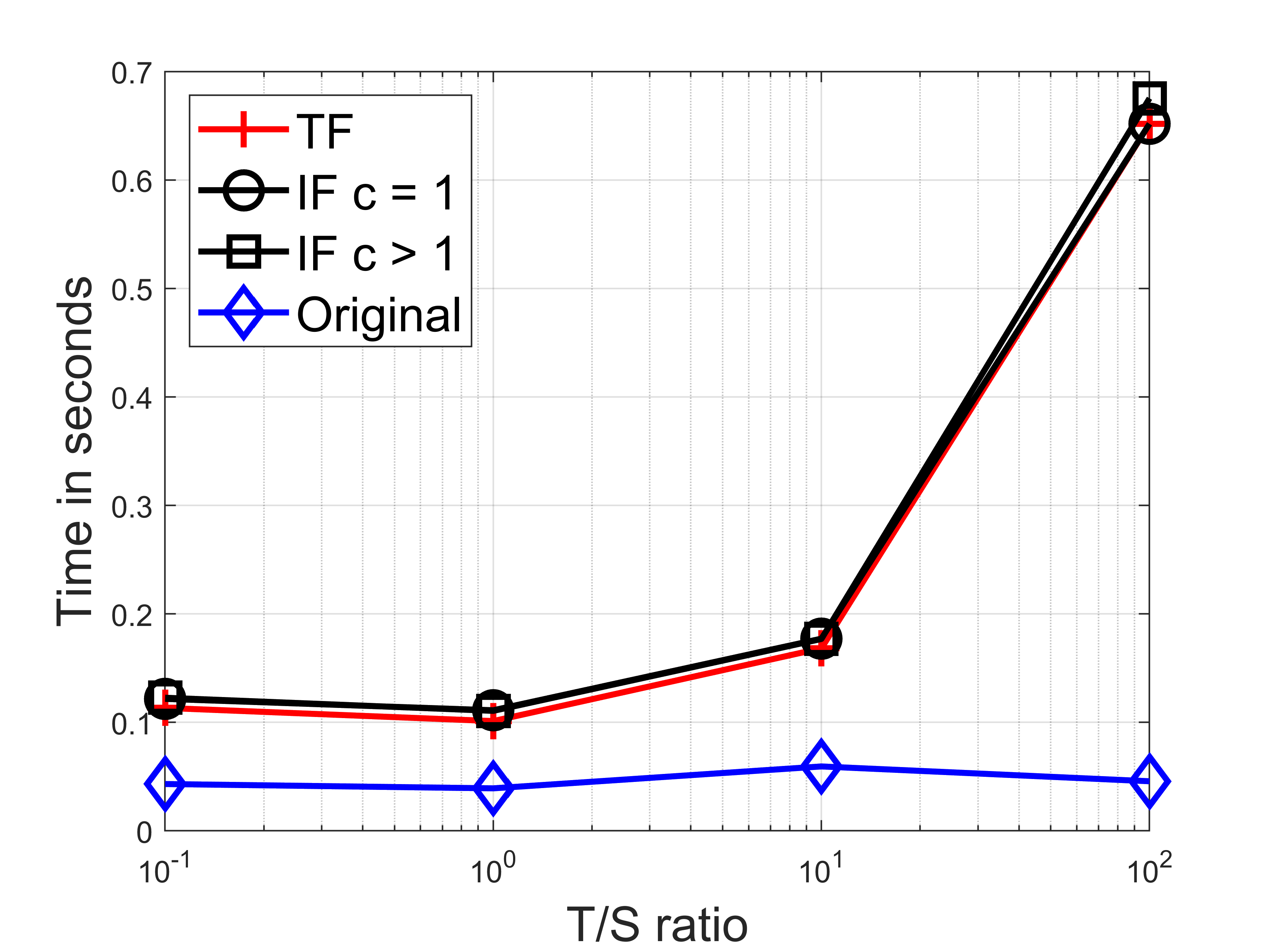}
  \includegraphics[scale=0.36]{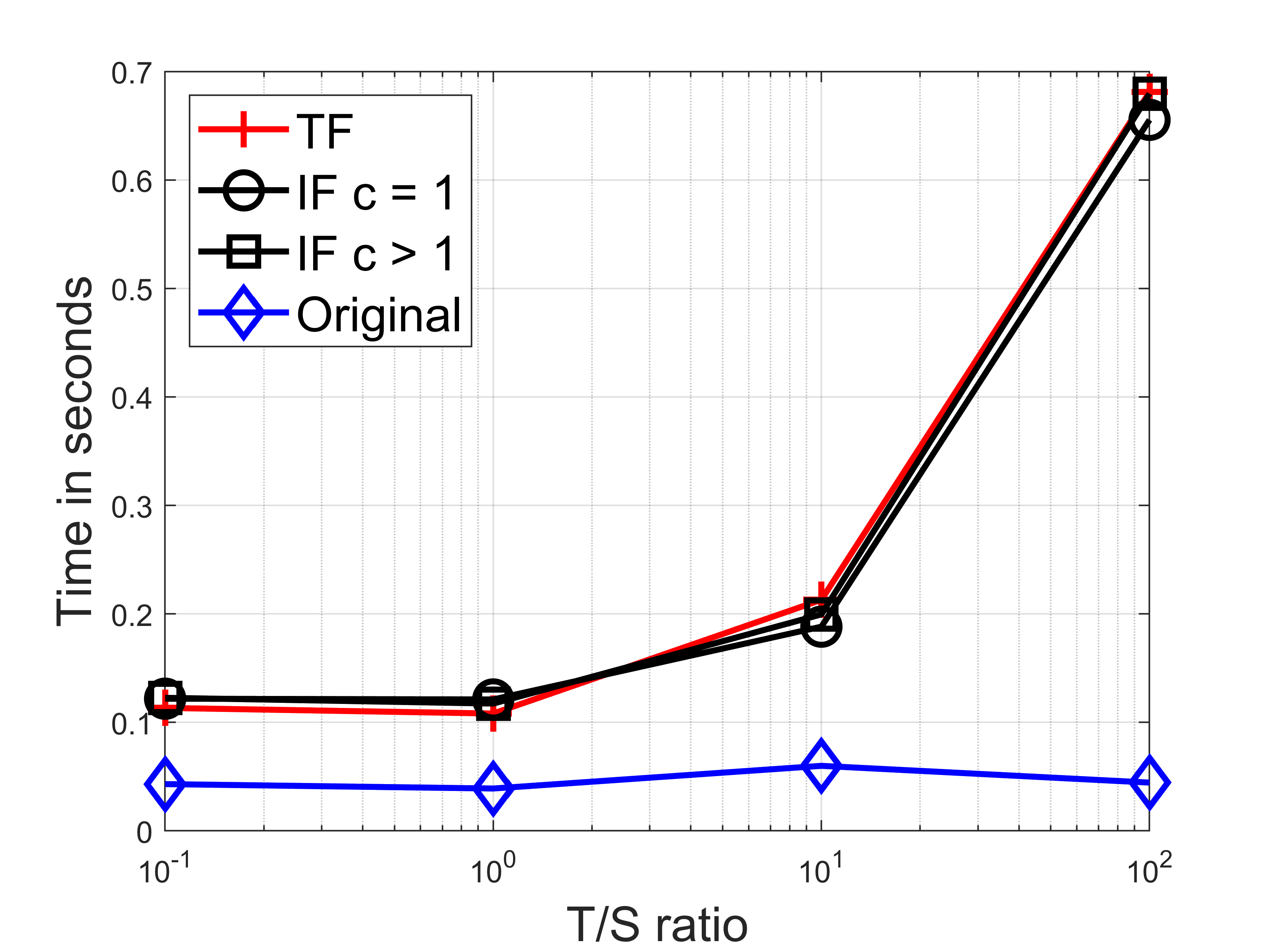}
  \caption{\small Average time needed to construct the filter when  $r=4$ (left), $r=8$ (middle), $r=16$ (right) and for $S$ of size one hundred thousand elements.  
  }\label{FigExp3CT}
\end{figure*}

The speed is measured for both for the construction of the filter and also for lookups. In the first case, the time needed to complete the construction was measured and compared to that of the original filter. The average over 1000 runs was computed and the results are shown in Figure \ref{FigExp3CT}. It can be seen that the proposed filters introduce a significant overhead that is around 2-3x when $T$ is small and increases with the size of $T$. This overhead is due to several  factors. The first one is the need to construct several filters, two for the TF and IF, $c=1$ constructions and in some cases even three for the IF, $c=2$ construction. The construction of the filters is also linked in the case of IF construction so that when trying a set of hash functions, construction is successful only when the construction of all the filters succeeds. The second factor is that some of those filters are larger as they store elements from $F$ in addition to those in $S$. The third factor that contributes to the overhead is the fact that all elements in $T$ have to be checked in the first filter to construct the set $F$. This means that as the size of $T$ increases, so does the overhead as it is clearly seen in the results on Figure \ref{FigExp3CT}. In any case, the construction time remains in the order of seconds even for large $T$. Additionally, the check for the elements in $T$ can be easily run in parallel if in some application the overhead becomes an issue. Indeed, in most applications construction is amortized over many query operations so its cost is not a major concern.

\begin{figure*}[h]
  \centering
  \includegraphics[scale=0.36]{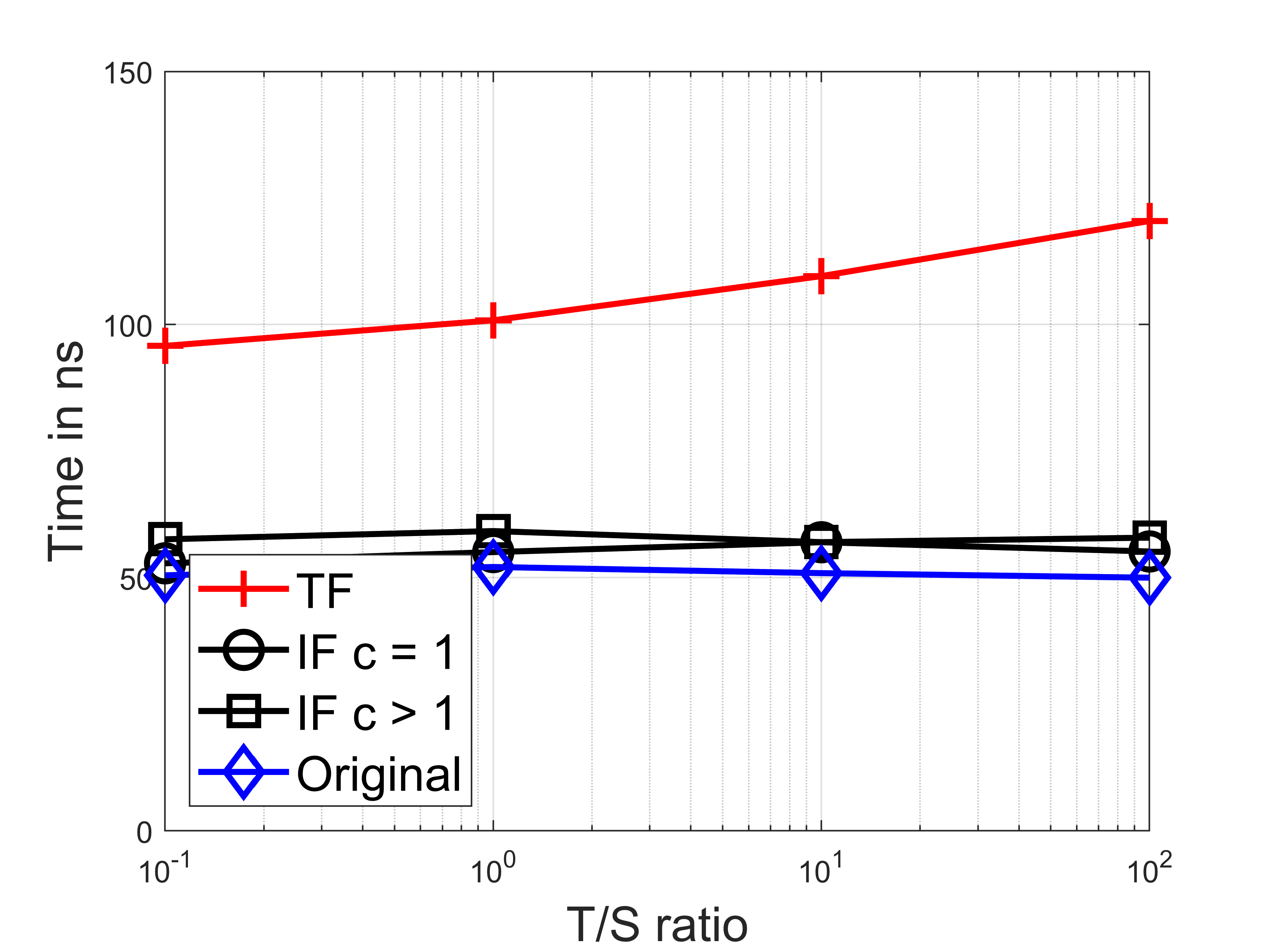}
  \includegraphics[scale=0.36]{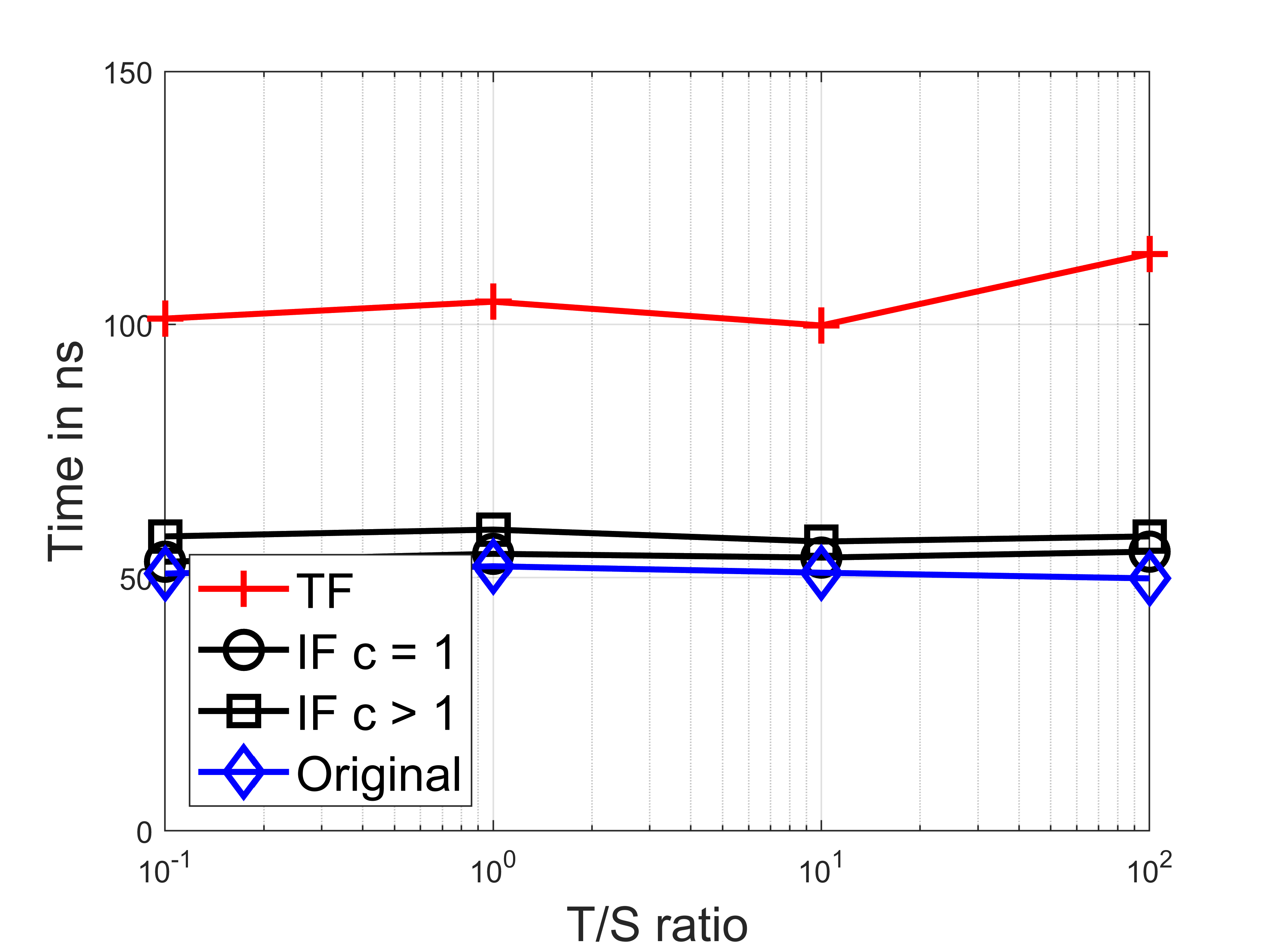}
  \includegraphics[scale=0.36]{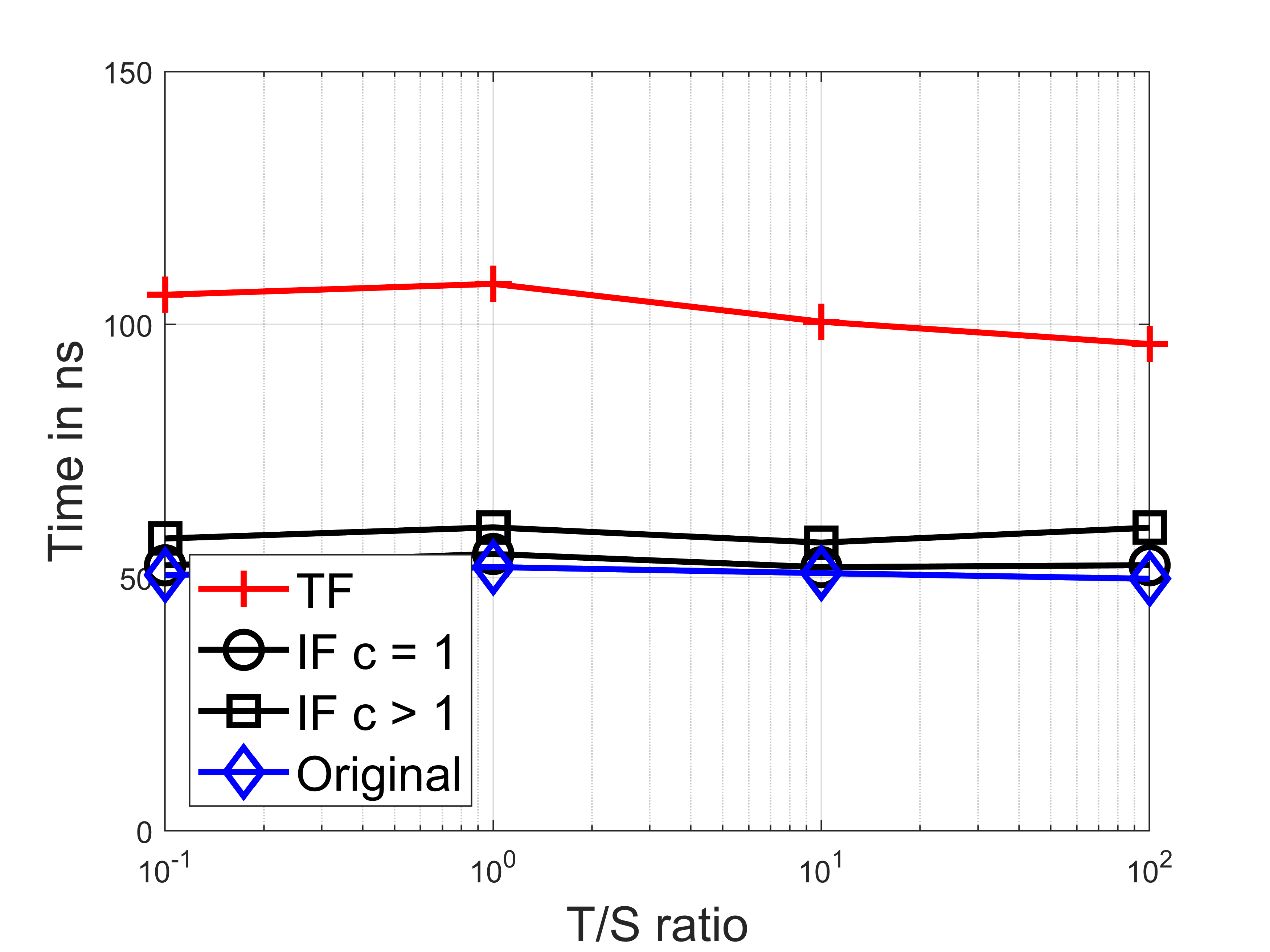}
  \caption{\small Average time needed for a positive lookup when  $r=4$ (left), $r=8$ (middle), $r=16$ (right) and for $S$ of size one hundred thousand elements.   
  }\label{FigExp3PL}
\end{figure*}

\begin{figure*}[h]
  \centering
  \includegraphics[scale=0.36]{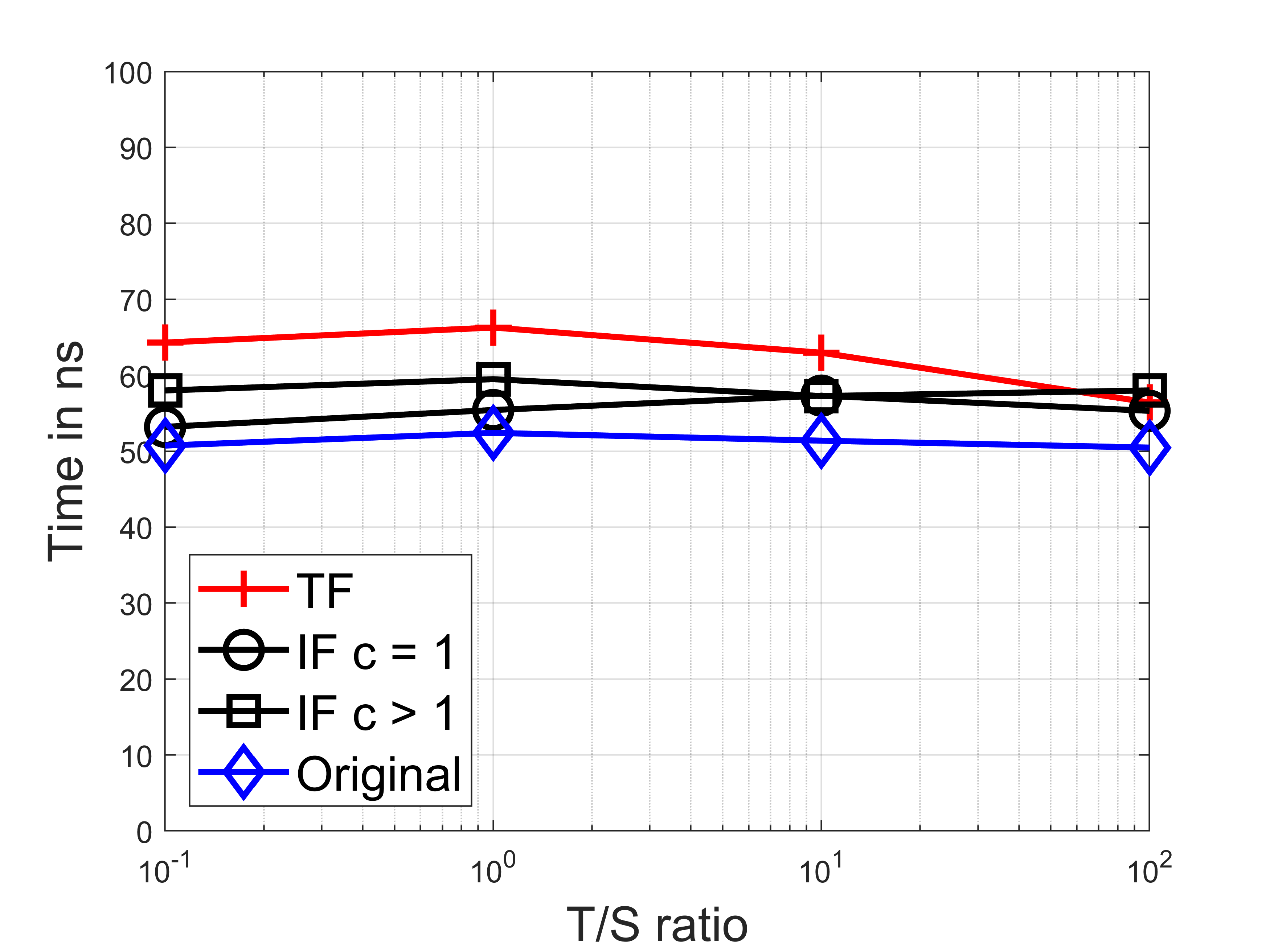}
  \includegraphics[scale=0.36]{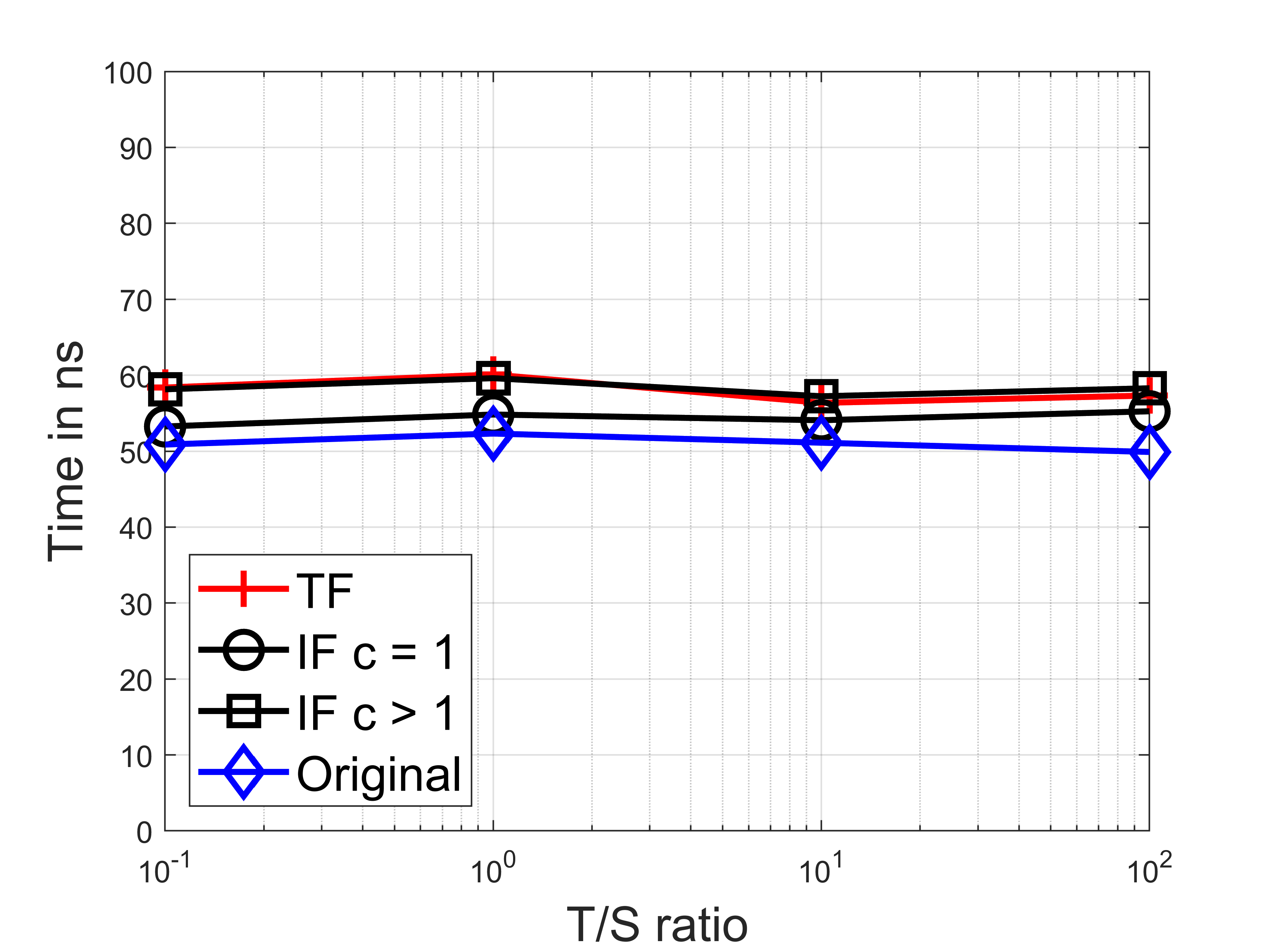}
  \includegraphics[scale=0.36]{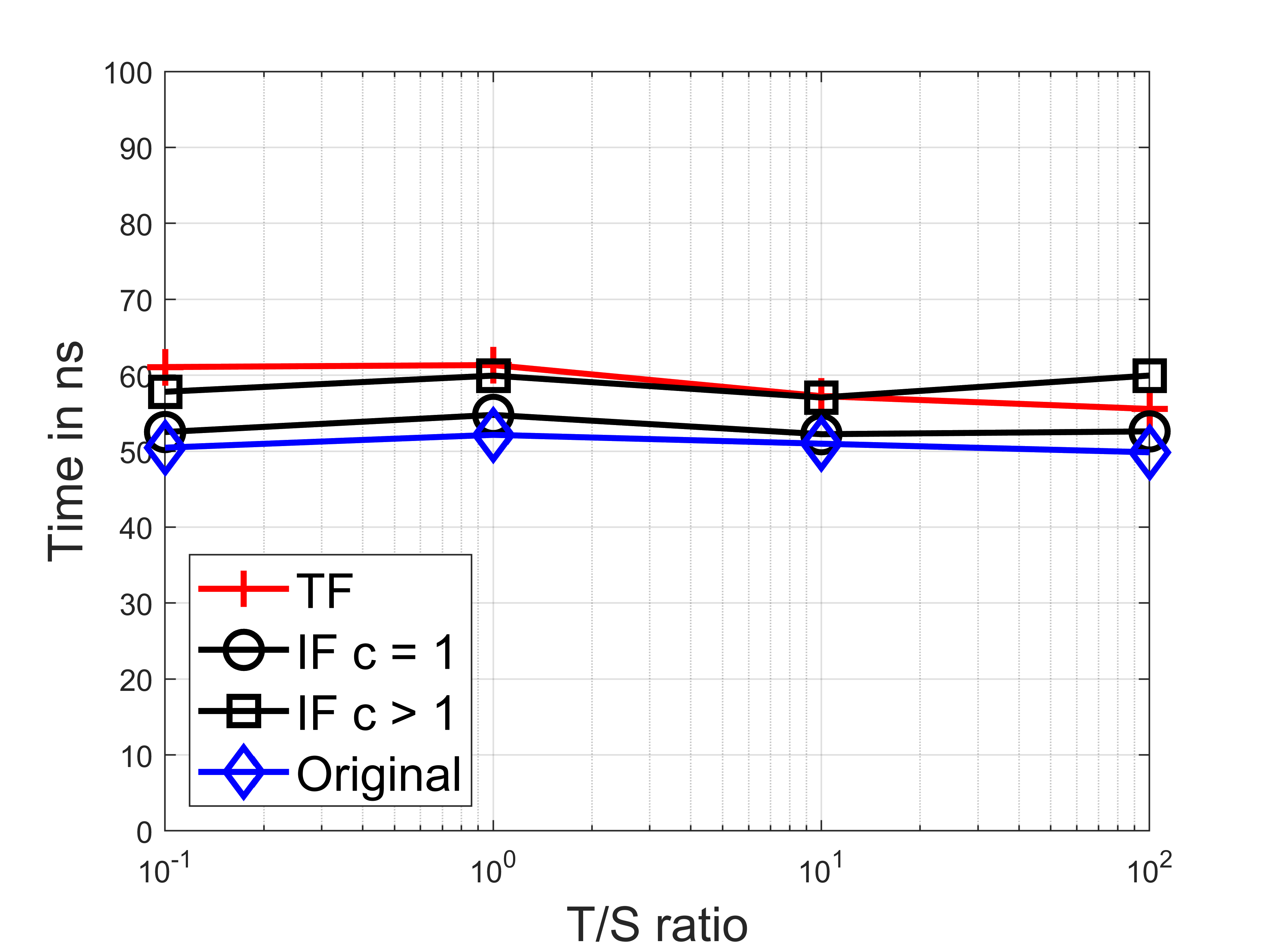}
\caption{\small Average time needed for a negative lookup when  $r=4$ (left), $r=8$ (middle), $r=16$ (right) and for $S$ of size one hundred thousand elements.    
  }\label{FigExp3NL}
\end{figure*}

To evaluate the speed of lookups we measure the speed for positives and negatives. Positives would require to check all the filters while for negatives most of them will be identified by the first filter avoiding access to the second so the overhead should be smaller. For positives, all the elements in $S$ were queried and the average time was logged while for negatives, one million random elements were tested. The results averaged over 1000 runs are shown in Figures \ref{FigExp3PL} and \ref{FigExp3NL} where the results for the original Xor filter are also included for comparison. It can be seen that the proposed filters introduce an overhead for positives. This lookup time is approximately 2x for the TF construction which needs to access different memory positions for each filter. Instead, for the IF constructions it is much lower, below 1.2x that of the original filter in all cases thus confirming the benefits of integrating the filters when positive lookup speed is a priority. Finally, speed does not have any clear dependency with the size of $T$ or the value of $r$ for positive lookups.

For negative lookups, the proposed filters also introduce a small overhead. In the case of the TF construction, the overhead is small as the first filter is the same and most lookups (approximately $1-2^{-r}$) only need to check the first filter. The IF constructions have also a small overhead because checking the first filter needs some additional operations to extract the relevant bits that share the word with those of the second filter. In any case, the lookup time is only approximately 1.2x that of the original filter in the worst case. In summary, the proposed filters introduce a significant overhead in the construction time while for lookups, the overhead is smaller and can be minimized for positive or negative lookups by selecting the filter construction. 

The implementation and evaluation confirm that false positive free on a set can be implemented with a moderate overhead in memory and speed when using the xor filter as the base structure. Therefore, for applications on which the cost of a false positive is large, the proposed filters can provide significant benefits by ensuring that the most frequent negative elements do not suffer false positives.

\section{Case studies}
\label{Cases}

This section discusses the applicability of the proposed filters by presenting a few use cases and showing the benefits of the proposed filters. The first case study is a simple spell checker, one of the original applications of Bloom filters, that is used to illustrate how the proposed filters work. Then, two more practical case studies a URL deny list and a cryptocurrency application are discussed.

\subsection{Spell checker}

One of the original applications of Bloom filters was to store the words of a dictionary to detect spelling errors \cite{BFSpellCheck}. Although this is no longer implemented with Bloom filters, it can serve to illustrate how our FPFS filters work. In this design, the checker fails to detect spelling errors that have a false positive on the filter. Unfortunately, those may occur for frequent spelling mistakes and it would be beneficial to avoid them. In this scenario, our filter with a false positive free sets can be used by selecting common spelling errors and adding them to set $T$. To illustrate this use case, we have taken the Birkbeck dataset\footnote{Available at https://www.dcs.bbk.ac.uk/\%7EROGER/corpora.html} that contains 36\,133 common misspellings of 6\,136 words and inserted the words\footnote{We have removed common misspellings that are duplicated or correspond to valid words so that the total number of misspelling is reduced to 32\,894.} in our filter with $r=8$ as set $S$ and the misspellings as set $T$. We verified that there are no false positives for those misspellings and measured the false positive rate for other words. The results show that the proposed filters require 60\,951, 63\,168 and 68\,229 memory bits for the TF, IF with $c=1$ and IF with $c=2$ constructions compared to 60\,624 memory bits of a traditional filter. The memory overhead is negligible (0.5\%) for the TF construction and very small (4.2\%) for the IF with $c=1$. The proposed filters achieve a similar false positive probability for negative elements not in $T$ as that of the original filter while eliminating all false positives in $T$. Instead, the original filter suffers on average 130.5 false positives on $T$, based on one thousand sample runs. This illustrates the benefits of the proposed filters in this simple case study.

\subsection{URL deny list} 

Another scenario where our filters can be useful is to implement a deny list of malicious URLs. A filter that stores the list of malicious URLs can be useful to perform an initial checking so that on a negative, the URL can be safely classified as non-malicious. Instead on a positive a full check would be needed to ensure that the URL is not a false positive. Here again, there are many valid URLs that are frequently used and thus ensuring that they do not suffer false positives would be beneficial. As an example, we have considered the URL dataset used in \cite{ADA-LBF} that has 485\,730 unique URLs of which 16.47\% are malicious, and the rest are benign. We store the malicious URLs in a filter using the benign URLs as set $T$. In this case, we set $r=8$ and the proposed filters require 656\,328, 679\,584 and 734\,697 memory bits for the TF, IF with $c=1$ and IF with $c=2$ constructions compared to 653\,040 memory bits of a traditional filter. Again, the memory overhead is negligible (0.5\%) for the TF construction and very small (4.1\%) for the IF with $c=1$. We verified that there are no false positives for the set $T$ in our filters while the traditional one suffers 1\,358 false positives on average, based on one thousand constructions. This means that again, we are able to avoid false positives on frequently used URLs. There are other similar use cases of access/deny lists where our filters could be useful, such as for IP addresses or person names (e.g. no-fly list).

\subsection{Bitcoin Simplified Payment Verification (SPVs) Nodes}

Bitcoin defines two modes of operation for the nodes: Full nodes and Simplified Payment Verification nodes (SPVs)\footnote{https://developer.bitcoin.org/devguide/operating\%5Fmodes.html}. The full nodes store all the blocks in the Bitcoin blockchain (which currently requires more than 300 GBs) while the SPV nodes only store the block headers. This reduces the storage and bandwidth needed by SPV nodes. Since SPV nodes do not have the full data of the transactions, they need to get it from the full nodes when needed. Originally, when an SPV needed to retrieve a transaction, it downloaded the blocks until the transaction was found which was very inefficient. To reduce the overhead, a first solution was implemented by the SPV nodes constructing a Bloom filter with the identifiers of the transactions that they want to get and sending it to the full nodes. Then, the full nodes applied the filter to each block and whenever a positive was returned by the filter for any of the transaction identifiers on the block, the block was sent to the SPV node that issued the request\footnote{https://github.com/bitcoin/bips/blob/master/bip-0037.mediawiki}. However, this scheme had privacy issues as full nodes can infer information from the SPV nodes \cite{privacyblockchainggervais2014} and the filters can also be used to launch denial of service attacks.  This led to a new design in which full nodes add a filter to each of the header blocks so that now SPV nodes can use those filters to locate the blocks that contain the transactions they are interested in. Once the blocks are identified, the SPV node can download only those blocks from the full nodes\footnote{https://github.com/bitcoin/bips/blob/master/bip-0157.mediawiki}. In this second version, Golomb Code Sets \cite{Golomb} are used instead of Bloom filters for the approximate membership checking\footnote{https://github.com/bitcoin/bips/blob/master/bip-0158.mediawiki}.

The number of transactions per month in Bitcoin is approximately 20 million and as of today the number of existing transactions is around 700 million\footnote{https://www.blockchain.com/charts/n-transactions-total}. Therefore, one possibility could be to replace Golomb Code Sets with our filter with a false positive free set using the universe of existing transactions as the set free of false positives or if that is not possible avoiding for example false positives for the transactions in the last few days that are more likely to be checked by the SPV nodes. As the number of transactions per block is around 2500, we have constructed filters with $S$ of size 2500 and $T$ of sizes 700 million and 20 million to implement a FPFS on all existing transactions and on the last 30 day transactions respectively. The FPFS filters when we set $r=8$, require in the first case (700 million) 62293 bits for the TF construction and 76245 and 65268 for the IF1 and IF2 constructions and in the second 46363 bits for the TF construction and 59262 and 46620 bits for the IF1 and IF2 constructions. This shows that the proposed filters can be used to completely avoid false positives on Bitcoin SPV nodes for all existing transactions or for the transactions done in the last 30 days. 

\section{Conclusions and Future Work}
\label{Conclusions}

In this paper, a new type of approximate membership query filters that completely eliminate false positives for a given set has been proposed. Several constructions based on xor filters have been presented and evaluated for these filters with a false positive free set showing that it is indeed possible to avoid false positives for large sets. The benefits of the filters with a false positive free set have also been illustrated with three practical case studies covering different applications in computing and networking. 

The constructions presented in this paper are also directly applicable to the ribbon filter \cite{Ribbonfilter}, which offers more design flexibility than standard xor filters. Therefore, implementing the proposed filter with a false positive free set using ribbon filters is an interesting topic to continue this work. Similarly, incorporating the proposed filters in Bitcoin SPV nodes and evaluating their benefits over existing filters would also be of interest. More broadly, we expect the proposed filters to be beneficial in many other scenarios and thus identifying those and evaluating the use of the proposed filters on them could lead to significant performance improvements. 

\section{Acknowledgements}

We would like to thank Mario Palomares for the discussions on the Bitcoin SPV case study. Pedro Reviriego would like to acknowledge the support of the ACHILLES project PID2019-104207RB-I00 and the Go2Edge network RED2018-102585-T funded by the Spanish Agencia Estatal de Investigaci\'on (AEI) 10.13039/501100011033 and of the Madrid Community research project TAPIR-CM grant no. P2018/TCS-4496. Stefan Walzer work was supported by the DFG grant WA 5025/1-1.

\balance
\bibliographystyle{IEEEtran}
\bibliography{BFrefs}

\end{document}